\documentclass[useAMS,usenatbib]{mnras}
\usepackage{graphicx,natbib,amssymb,amsmath,stmaryrd,makecell}
\usepackage{txfonts}
\usepackage{xcolor}
\usepackage{hyperref}
\usepackage{xspace,dsfont}
\usepackage{caption, lipsum, comment}

\graphicspath{{Figures/}}

\newcommand{\Kel}{{\rm K}}

\newcommand{\expf}[1]{{{\rm e}^{#1}}}

\newcommand{\vgh}{{\hat{\boldsymbol\gamma}}}
\newcommand{\vghp}{{\hat{\boldsymbol\gamma}'}}

\newcommand{\id}{{\,\rm d}}

\newcommand{\beq}{\begin{equation}}   %

\newcommand{\eeq}{\end{equation}}   %

\newcommand{\beqa}{\begin{eqnarray}}   %

\newcommand{\eeqa}{\end{eqnarray}}   %

\newcommand{\bealf}[1]{\begin{align} #1 \end{align}}

\newcommand{\beal}{\begin{align}}
\newcommand{\enal}{\end{align}}

\newcommand{\bspl}{\begin{split}}

\newcommand{\espl}{\end{split}}

\newcommand{\bsub}{\begin{subequations}}

\newcommand{\esub}{\end{subequations}}

\newcommand{\bmulti}{\begin{multline}}   %

\newcommand{\beqm}{\begin{mathletters}}   %

\newcommand{\eeqm}{\end{mathletters}}   %

\newcommand{\Ne}{N_{\rm e}}

\newcommand{\sigT}{\sigma_{\rm T}}

\newcommand{\vek} [1]{\mbox{\boldmath${#1}$\unboldmath}}

\newcommand{\pot}[2]{#1 \times 10^{#2}}

\newcommand{\oDnu}{{\mathcal{\hat{D}}_{\nu}}}

\newcommand{\oOnu}{{\mathcal{\hat{O}}_{\nu}}}

\renewcommand{\Ne}{N_{\rm e}}

\usepackage{comment}

\newcommand{\threej}[6]{ \begin{pmatrix}
    #1 & #2 & #3 \\
    #4 & #5 & #6
\end{pmatrix}}

\title[Photon Boost Operator]
{The Boost Operator: Properties, Computation and Applications}

\author[Chluba and Ravenni et al.]{
Jens~Chluba$^{1}$\thanks{E-mail:Jens.Chluba@Manchester.ac.uk} and Andrea Ravenni$^{1}$
\\
$^{1}$Jodrell Bank Centre for Astrophysics, Alan Turing Building, University of Manchester, Manchester M13 9PL
}

\begin{document}

\date{\vspace{-5mm}{Accepted 2021 --. Received 2021}}

\maketitle

\begin{abstract}
The transformation of radiation signals (e.g., photon occupation number and integrated intensity) between moving frames is a common task is physics, astrophysics and cosmology. Here we show that the required {\it boost operator}, relating the frequency-dependent spin-weighted spherical harmonic coefficients of the considered observable between the frames, is directly given by the {\it aberration kernel} with the Doppler weight parameter being replaced by a differential operator. The aberration kernel has been previously studied in great detail, meaning that this simplification allows us to directly compute the boost operator using the expressions of the aberration kernel. 
As a preparatory step, we generalize the differential equation that determines the aberration kernel to general Doppler weight. This avoids the intermediate step of Doppler-weight raising and lowering operations in computations of the boost operator. We then clarify all the properties of the boost operator (e.g., raising and lowering operations, symmetries and commutation relations) and derive a formal operator differential equation for the boost operator. This differential equation allows us to quickly generate the boost operator for which we give exact expressions up to second order in $\varv/c$. 
For illustration, we then apply the boost operator to transformations of the cosmic microwave background (CMB), validating that measurements of the lowest CMB multipoles do not allow determining the amplitude of the primordial CMB dipole. We also derive the kinematic corrections to the Thomson scattering process (to all orders in $\varv/c$), giving explicit expressions up to second order in $\varv/c$, showcasing an application of the boost operator in radiative transfer problems.
\end{abstract}


\section{Introduction}
\label{sec:Intro}
In the modeling of physical interactions and measurements we often have to relate radiation observables in one frame of reference to the quantities of another that is moving with a relative speed $\beta=\varv/c$. The two frames are related by Lorentz boosts, in principle only requiring straightforward manipulations in the transformation. However, in detail the computations can become quite complicated and furthermore have to be repeated many times, begging the question if a general approach can be developed that is transparent and avoids the repetitive aspects of the calculation. 

In many applications, we consider photon signals across the sky, which we can decompose using spin-weighted spherical harmonics, ${_s}Y_{\ell m}(\theta, \phi)$. For observables that only depend on the observation direction, $X(\theta, \phi)$ [e.g., the thermodynamic temperature of the cosmic microwave background (CMB)], the {\it aberration kernel} provides a simple way to relate the spherical harmonic coefficients in one frame to those of another \citep[e.g.][for related theory papers]{Challinor2002, Chluba2011ab, Dai2014}, giving the transformation law
$$X'_{\ell m}=\sum_{\ell', m'} {_s^d}\mathcal{K}^{mm'}_{\ell \ell'}(\beta)\,X_{\ell' m'},$$
where $X_{\ell m}$ gives the spherical harmonic coefficients in the resting frame and $X'_{\ell m}$ those in the moving frame. The aberration kernel, ${_s^d}\mathcal{K}^{mm'}_{\ell \ell'}(\beta)$, describes the motion-induced mixing between the angular momentum, $\ell$, and magnetic quantum numbers, $m$, while the the spin-weight, $s$, is unaffected. 
For $\beta=0$, the kernel reduces to the identity, ${_s^d}\mathcal{K}^{mm'}_{\ell \ell'}(\beta=0)=\delta_{\ell\ell'}\delta_{mm'}$. For $\boldsymbol{\beta}$ parallel to the $z$-axis\footnote{The general case can be obtained by rotating the coordinate system \citep[e.g.,][]{Challinor2002, Dai2014}.}, no mixing between magnetic quantum numbers occurs, even if the aberration kernel explicitly depends $m$, i.e., ${_s^d}\mathcal{K}^{mm'}_{\ell \ell'}(\beta)\rightarrow {_s^d}\mathcal{K}^{m}_{\ell \ell'}(\beta)\,\delta_{mm'}$. In addition, the Doppler weight\footnote{This essentially counts the powers of photon frequency appearing in the Lorentz invariance property of the observable. For instance, thermodynamic CMB temperature variables have Doppler weight $d=1$, photon occupation number has $d=0$ and specific intensity has Doppler weight $d=3$. } of the observable, $d$, influences the exact coupling matrix elements. 
By pre-computing the kernel, the transformation problem for fields $X(\theta, \phi)$ is therefore solved.

Explicit calculations of the aberration kernel become challenging for large $\ell$. For $\beta\ll 1$, Taylor series expansions were developed up to finite order in $\beta$ \citep[e.g.,][]{Challinor2002, Amendola2010, Chluba2011ab}. Recurrence relations open a principal way to compute the aberration kernel to any order \citep{Chluba2011ab, Dai2014}; however, currently the most efficient and numerically stable way is based on the kernel differential equation for $d=1$ \citep{Dai2014, Yasini2017}. The kernel for other Doppler weights can then be obtained using Doppler-weight raising and lowering operations \citep{Dai2014, Yasini2017}. In Sect.~\ref{sec:K_ODE} we show that this can be simplified by generalizing the kernel differential equation to general Doppler weight. 

Although for observables $X(\theta, \phi)$ the general transformation problem is solved, for observables that also depend on the photon frequency, $X(\nu, \theta, \phi)$, some work is left. The final transformation law can be expressed as 
$$X'_{\ell m}(\nu)=\sum_{\ell', m'} {_s^d}\hat{\mathcal{B}}^{mm'}_{\ell \ell'}(\nu, \beta)\,X_{\ell' m'}(\nu),$$
which introduces the {\it boost operator}\footnote{A series expansion in $\nu$ is used to model the transformation of each derivative spectrum, $\nu^k \partial_\nu^k \, X_{\ell m}(\nu)$ of the observable individually [e.g., see Eq.~\eqref{eq:n_mov}].}, ${_s^d}\hat{\mathcal{B}}^{mm'}_{\ell \ell'}(\nu, \beta)$. Once pre-computed, this again solves the transformation problem.

In \citet{Challinor2002}, expressions describing the {\it observer}\footnote{This is simply related to the sign convention of $\boldsymbol{\beta}$.} boost operator up to second order in $\beta$ were considered. This was analytically generalized for to all orders in $\beta$ by using the aberration kernel \citep{Yasini2017}. However, the resulting boost operator expression requires the computation of the aberration kernel for infinitely many Doppler weights. In addition, the properties of the boost operator were not considered in detail. 

In this work, we show that formally the boost operator is simply given by\footnote{Here we again assume $\boldsymbol{\beta}$ parallel to the $z$-axis noting that the general case can be obtained by rotation.} 
$${_s^d}\hat{\mathcal{B}}^{m}_{\ell \ell'}(\nu, \beta)\equiv {_s^{d+\oOnu}}\mathcal{K}^{m}_{\ell \ell'}(-\beta),$$
where $\oOnu=-\nu\partial_\nu$ is the {\it energy shift generator}. This means that we can directly use all the relations of the aberration kernel in calculations of the boost operator when treating the Doppler weight as an operator, $d+\oOnu$, greatly simplifying matters. We furthermore obtain a simple operator differential equation for ${_s^d}\hat{\mathcal{B}}^{m}_{\ell \ell'}(\nu, \beta)$ that can be used to compute the boost operator term-by-term. 

With these new insights, it is easy to apply the boost operator to physical problems. In Sect.~\ref{sec:dipole_quadrupole}, we apply the boost operator formalism to the transformation of the low-$\ell$ CMB anisotropies in the presence of a primordial quadrupole. This illustration validates that one cannot extract the primordial dipole from measurements of the CMB monopole, dipole and quadrupole alone \citep[see also][]{Kamionkowski2003}. In Sect.~\ref{sec:kin_corrs_Thomson}, we use the boost operator formalism to compute the kinematic corrections to Thomson scattering, obtaining a formal solution to all orders in $\beta$ and explicit expressions up to second order. These examples illustrate how useful the boost operator approach can be in astrophysical problems.

\section{The boost operator and its properties}
\label{sec:boost_operator}
In this section, we develop a basic understanding of the boost operator approach. We briefly recap some of the previous steps taken in the literature and then study the properties and ways to compute the boost operator in more detail. Parts of the proofs will be detailed in Appendices to keep the presentation tidy.

\subsection{Lorentz transformations using the aberration kernel}
\label{sec:CS_lorentz_abb}
The main task is to understand how various quantities that are described using spherical harmonics\footnote{For now, we focus on spin-weight $s=0$, but will generalize below.}, $Y_{\ell m}$, transform under Lorentz boosts. In many problem, a key quantity is the photon occupation number, $n(\nu, \vgh)$, for photons traveling along the direction $\vgh$ at frequency $\nu$. The relevant transformation into the moving frame can be obtained using the aberration kernel \citep[e.g.,][]{Challinor2002, Dai2014} and its generalized form \citep{Yasini2017}. Denoting variables in the moving frame using primes, we may write\footnote{We suppress the explicit dependence on time $t$ and position $\vek{x}$ in space, which are not affected for time-independent observables.}
\bealf{
\label{eq:n_mov}
n'(\nu', \vghp)&=n(\nu, \vgh)\equiv \sum_{k=0}^\infty \frac{(\nu-\nu')^k}{k!} \frac{\partial^k}{\partial {\nu'}^k} n(\nu', \vgh) 
\nonumber\\
&=\sum_{\ell m}  \sum_{k=0}^\infty \left[\frac{(\nu-\nu')^k}{{\nu'}^k}  Y_{\ell m}(\vgh) \right] \,\frac{{\nu'}^k}{k!}  \frac{\partial^k}{\partial {\nu'}^k} n_{\ell m}(\nu'),
}
where $\nu=\nu(\nu', \vghp)$ and $\vgh=\vgh(\vghp)$ can be expressed in terms of the moving frame coordinates. For the first equality in Eq.~\eqref{eq:n_mov}, we used the invariance of the photon phase space distribution. For the second equality, we performed a Taylor series around $\nu'$, and for the third equality we also added the spherical harmonic expansion of $n(\nu', \vgh)$ in the lab frame evaluated at $\vgh=\vgh(\vghp)$.

For simplicity, we shall align the direction of the motion (or boost) with the positive $z$-axis assuming a speed $\beta=\varv/c$. With this choice, only mixing between the angular momentum quantum numbers $\ell$ occurs. The general case can be obtained by rotating the final expression \citep[e.g.,][]{Dai2014}. Using ${\boldsymbol \beta}\parallel {\boldsymbol z}$ directly implies $\phi'=\phi$ and the transformation laws 
\bsub
\bealf{
\nu'&=\gamma \nu(1-\beta \mu), \qquad \mu'=\frac{\mu-\beta}{1-\beta \mu}
\\
\nu&=\gamma \nu'(1+\beta \mu'), \qquad \mu=\frac{\mu'+\beta}{1+\beta \mu'}
}
\esub
with Lorentz factor $\gamma=(1-\beta^2)^{-1/2}$ and direction cosines $\mu=\cos\theta$ and $\mu'=\cos\theta'$.

Next we calculate the transformation law of $(\nu/\nu'-1)^k  Y_{\ell m}(\vgh)$, which can be expressed as
\bealf{
\label{eq:def_H}
^{k}\mathcal{H}_{\ell\ell'}^{m}(\beta)
&= \int  Y^*_{\ell m}(\vghp) \left[\frac{(\nu-\nu')^k}{{\nu'}^k} \right]  Y_{\ell' m}(\vgh) \id\vghp
\nonumber\\
&= \sum_{t=0}^k (-1)^{k-t}\,\binom {\,k\,} {\,t\,}\;
\int  Y^*_{\ell m}(\vghp) \left[\frac{\nu}{\nu'} \right]^t  Y_{\ell' m}(\vgh) \id\vghp
\nonumber\\
&=\sum_{t=0}^k \; (-1)^{k-t}\,\binom {\,k\,} {\,t\,}\; ^{-t}\mathcal{K}_{\ell\ell'}^{m}(-\beta).
}
Here, $^{d}\mathcal{K}_{\ell\ell'}^{m}(\beta)$ is the aberration kernel of Doppler weight $d$, which is explicitly defined as \citep[e.g.,][]{Dai2014}:
\bealf{
\label{eq:K_abb_def}
^{d}\mathcal{K}_{\ell\ell'}^{m}(\beta)
&= \int  \frac{Y^*_{\ell m}(\vgh') Y_{\ell' m}(\vgh)}{[\gamma(1-\beta \mu')]^d} \id\vghp
}
and can be computed exactly.
It is important to mention that for $^{k}\mathcal{H}_{\ell\ell'}^{m}(\beta)$ we need $^{d}\mathcal{K}_{\ell\ell'}^{m}(\beta)$ for {\it negative} $\beta$, as the aberration kernel was originally defined as an {\it observer} kernel. When observing a photon in a direction $\vgh$, it propagates along the direction $-\vgh$, explaining this difference \citep[see][]{Challinor2002}.
Note also that because in Eq.~\eqref{eq:n_mov} powers of $\nu/\nu'=\gamma(1+\beta \mu')$ appear, here we need the aberration kernel for Doppler weights $d\leq 0$. 

Overall, we then have the transformed spherical harmonic coefficient of the occupation number in the moving frame as
\bealf{
\label{eq:n_mov_final}
n'_{\ell m}(\nu')&=\sum_{\ell'} \sum_{k=0}^\infty
{^{k}\mathcal{H}_{\ell\ell '}^{m}(\beta)}\,\frac{{\nu'}^k}{k!}  \frac{\partial^k}{\partial {\nu'}^k} n_{\ell' m}(\nu')
\nonumber\\
&\equiv \sum_{\ell'} \mathcal{\hat{B}}_{\ell\ell '}^{m}(\nu', \beta) \,n_{\ell' m}(\nu'),
}
where in the second line we introduced the {\it boost operator}
\bealf{
\label{eq:gen_kernel_op}
\mathcal{\hat{B}}_{\ell\ell'}^{m}(\nu, \beta)
&\equiv
\sum_{k=0}^\infty
 \int  Y^*_{\ell m}(\vghp) \left[\frac{(\nu-\nu')^k}{{\nu'}^k} \,\frac{{\nu}^k}{k!}  \frac{\partial^k}{\partial {\nu}^k} \right]  Y_{\ell' m}(\vgh) \id\vghp
\nonumber
\\
&=
\sum_{k=0}^\infty \sum_{t=0}^k \; (-1)^{k-t}\,\binom {\,k\,} {\,t\,}\; ^{-t}\mathcal{K}_{\ell\ell'}^{m}(-\beta)
\,\frac{{\nu}^k}{k!}  \frac{\partial^k}{\partial {\nu}^k}
\nonumber\\
&=\sum_{k=0}^\infty
{^{k}\mathcal{H}_{\ell\ell '}^{m}(\beta)}\,\frac{{\nu}^k}{k!}  \frac{\partial^k}{\partial {\nu}^k},
}
which identifies with the generalized aberration kernel operator for (starting) Doppler weight $d=0$ as previously discussed by \citet{Yasini2017} and defined for an observer (i.e., $-\beta$). We note that the aberration kernel is needed to infinite orders in the Doppler weight, an aspect that can be avoided, as we show below.

%

Before we venture into the details, we first generalize the boost operator to any {\it starting} Doppler weight $d$ and spin-weight. For the photon occupation number, the starting Doppler weight is $d=0$ and spin-weight $s=0$. However, for applications to  kinetic equations and also when treating more general quantities (e.g., like intensity and polarization states), the starting Doppler and spin weights may vary. This results in the generalization
\bealf{
\label{eq:gen_kernel_op_ds}
{_s^d}\mathcal{\hat{B}}_{\ell\ell'}^{m}(\nu, \beta)
&\equiv
\sum_{k=0}^\infty
\! \int  \!{_{-s}}Y^*_{\ell m}(\vghp) 
\left[
\left(\frac{\nu}{\nu'}\right)^{-d}\, 
\frac{(\nu-\nu')^k}{{\nu'}^k} \,
 \frac{{\nu}^k}{k!} \frac{\partial^k}{\partial {\nu}^k} \right]  
{_{-s}}Y_{\ell' m}(\vgh) \id\vghp
\nonumber
\\
&=
\sum_{k=0}^\infty \sum_{t=0}^k \; (-1)^{k-t}\,\binom {\,k\,} {\,t\,}\; {_s^{d-t}}\mathcal{K}_{\ell\ell'}^{m}(-\beta)
\,\frac{{\nu}^k}{k!}  \frac{\partial^k}{\partial {\nu}^k}
\nonumber\\
&=\sum_{k=0}^\infty
{_{s}^{d, k}\mathcal{H}_{\ell\ell '}^{m}(\beta)}\,\frac{{\nu}^k}{k!}  \frac{\partial^k}{\partial {\nu}^k},
}
which was also considered for the generalized (observer) aberration kernel \citep[see][]{Yasini2017}.
The above equation implicitly defines the generalized version of Eq.~\eqref{eq:def_H}
\bealf{
\label{eq:def_H_gen}
{_{s}^{d, k}\mathcal{H}_{\ell\ell '}^{m}(\beta)}
&\equiv
\int  \!{_{-s}}Y^*_{\ell m}(\vghp) 
\left[
\left(\frac{\nu}{\nu'}\right)^{-d}\, 
\frac{(\nu-\nu')^k}{{\nu'}^k}  \right]  
{_{-s}}Y_{\ell' m}(\vgh) \id\vghp
\nonumber
\\
&=
\sum_{t=0}^k \; (-1)^{k-t}\,\binom {\,k\,} {\,t\,}\; {_s^{d-t}}\mathcal{K}_{\ell\ell'}^{m}(-\beta).
}
Since both ${_s^d}\mathcal{\hat{B}}_{\ell\ell'}^{m}(\nu, \beta)$ and ${_{s}^{d, k}\mathcal{H}_{\ell\ell '}^{m}(\beta)}$ are directly related to the aberration kernel ${_s^{d-t}}\mathcal{K}_{\ell\ell'}^{m}(-\beta)$, it is easy to deduce several of their important properties, as we discuss next.

\subsubsection{Alternative version using the operator $\hat{\mathcal{O}}_\nu=-\nu \partial_\nu$}
\label{sec:Alternative_Onu}
Instead of performing the Taylor series in $\nu$, we can also use $\ln\nu$ as variable, which yields
\bealf{
\label{eq:gen_kernel_op_ds_o_nu}
{_s^d}\mathcal{\hat{B}}_{\ell\ell'}^{m}(\nu, \beta)
&\equiv
\sum_{k=0}^\infty
\! \int  \!{_{-s}}Y^*_{\ell m}(\vghp) 
\left[
\left(\frac{\nu}{\nu'}\right)^{-d} 
\frac{\ln(\nu/\nu')^k \,(-1)^k\hat{\mathcal{O}}_\nu^k}{k!}  \right]  
\!{_{-s}}Y_{\ell' m}(\vgh) \id\vghp
\nonumber
\\
&=
\int  \!{_{-s}}Y^*_{\ell m}(\vghp) 
\left[
\left(\frac{\nu'}{\nu}\right)^{d} 
\sum_{k=0}^\infty
\frac{\ln(\nu'/\nu)^k \,\hat{\mathcal{O}}_\nu^k}{k!}  \right] 
{_{-s}}Y_{\ell' m}(\vgh) \id\vghp
\nonumber
\\
&=
\int  \!{_{-s}}Y^*_{\ell m}(\vghp) 
\left\{
\left(\frac{\nu'}{\nu}\right)^{d}\, 
\exp\left[\ln(\nu'/\nu)\,\hat{\mathcal{O}}_\nu \right]  \right\}  
{_{-s}}Y_{\ell' m}(\vgh) \id\vghp
\nonumber
\\
&=
\int  \!{_{-s}}Y^*_{\ell m}(\vghp) 
\left[
\left(\frac{\nu'}{\nu}\right)^{d+\hat{\mathcal{O}}_\nu}\right] 
{_{-s}}Y_{\ell' m}(\vgh) \id\vghp
\nonumber
\\
&\equiv  {_s^{d+\hat{\mathcal{O}}_\nu}}\mathcal{K}_{\ell\ell'}^{m}(-\beta).
}
Here, we used that $\nu/\nu'$ is a function of the angles only, implying that it commutes with the operator $\hat{\mathcal{O}}_\nu=-\nu\partial_\nu$.

While the last equivalence in Eq.~\eqref{eq:gen_kernel_op_ds} is not directly applicable in numerical evaluations, it identifies a significant simplification of the problem: We can  obtain all analytic expressions for the boost operator using those for the aberration kernel only. Formally, the operator $\hat{\mathcal{O}}_\nu$ plays the role of a Doppler weight, allowing the replacement $d\rightarrow d+\hat{\mathcal{O}}_\nu$ in all aberration kernel expressions given by \citet{Dai2014}. This also yields a formal differential equation for the boost operator as we show in Sect.~\ref{sec:ODE_for_B}.

We also mention that $\frac{\beta\,\hat{\mathcal{O}}_\nu}{\sqrt{3}}\approx {_0^0}\mathcal{\hat{B}}_{10}^{0}(\nu, -\beta)$ is the observer boost operator at lowest order in $\beta$. It determines how the photon occupation number of the monopole leaks into the dipole of the radiation field. The same operator was used in \citet{chluba_spectro-spatial_2023-I} to reformulate the Kompaneets equation in terms of single Doppler boosts. We will see below that this operator reappears in radiative transfer problems and for the transformation of the CMB into the moving frame.

\subsection{Main properties of the boost operator}
\label{sec:properties_boost_op}
Given that the boost operator $_s^d\mathcal{\hat{B}}_{\ell\ell '}^{m}(\nu, \beta)$ is a linear combination of the aberration kernel, we can immediately read off some of its main properties from those of ${_s^{d}\mathcal{K}_{\ell\ell'}^{m}(\beta)}$ \citep[see][]{Dai2014}
\bsub
\label{eq:prop_K}
\bealf{
\label{eq:prop_K_a}
{_s^{d}\mathcal{K}_{\ell'\ell}^{m}(\beta)}&= {_{s}^{2-d}\mathcal{K}_{\ell\ell'}^{m}(-\beta)}=(-1)^{\ell+\ell'}\,{_{-s}^{2-d}\mathcal{K}_{\ell\ell'}^{m}(\beta)}
\\
\label{eq:prop_K_b}
{_s^{d}\mathcal{K}_{\ell\ell'}^{m}(-\beta)}&=(-1)^{\ell+\ell'}\,{_{-s}^{d}\mathcal{K}_{\ell\ell'}^{m}(\beta)}
\\
\label{eq:prop_K_c}
{_{-s}^{d}\mathcal{K}_{\ell\ell'}^{-m}(\beta)}&=\left[{_s^{d}\mathcal{K}_{\ell\ell'}^{m}(\beta)}\right]^*={_s^{d}\mathcal{K}_{\ell\ell'}^{m}(\beta)}.
}
\esub
Referring to Eq.~\eqref{eq:gen_kernel_op_ds} it is easy to understand that Eq.~\eqref{eq:prop_K_b} and \eqref{eq:prop_K_c} carry over trivially:
\bsub
\label{eq:prop_B}
\bealf{
{_s^{d}\mathcal{\hat{B}}_{\ell\ell'}^{m}(\nu, -\beta)}&=(-1)^{\ell+\ell'}\,{_{-s}^{d}\mathcal{\hat{B}}_{\ell\ell'}^{m}(\nu, \beta)}
\\
{_{-s}^{d}\mathcal{\hat{B}}_{\ell\ell'}^{-m}(\nu, \beta)}&=\left[{_s^{d}\mathcal{\hat{B}}_{\ell\ell'}^{m}(\nu, \beta)}\right]^*={_s^{d}\mathcal{\hat{B}}_{\ell\ell'}^{m}(\nu, \beta)}.
}
\esub
However, there is no simple equivalent of Eq.~\eqref{eq:prop_K_a} for switching $\ell$ and $\ell'$. 
By using Eq.~\eqref{eq:gen_kernel_op_ds_o_nu} instead of Eq.~\eqref{eq:gen_kernel_op_ds} we can formally show
\bealf{
\label{eq:gen_kernel_op_ds_o_nu_switch}
{_s^d}\mathcal{\hat{B}}_{\ell'\ell}^{m}(\nu, \beta)
&\equiv  {_s^{d+\hat{\mathcal{O}}_\nu}}\mathcal{K}_{\ell'\ell}^{m}(-\beta)
={_s^{2-d-\hat{\mathcal{O}}_\nu}}\mathcal{K}_{\ell\ell'}^{m}(\beta).
}
This expression can again be helpful in analytic derivations.
From similar considerations, Eq.~\eqref{eq:prop_B} carries over to ${_{s}^{d, k}\mathcal{H}_{\ell\ell '}^{m}(\beta)}$ but no simple $\ell$-switching procedure was found.

Since $(\nu/\nu'-1)^k$ is a polynomial in $\gamma \beta \mu'$, we also know that up to a fixed order $N_\beta$ in $\beta$ the sum in ${^d_s} \mathcal{\hat{B}}_{\ell\ell '}^{m}(\nu, \beta)$ remains finite with $k\leq N_\beta$, thus limiting the frequency derivatives. For example, up to second order in $\beta$ we only need the aberration kernels for three Doppler weights $d$, $d-1$ and $d-2$ giving at most terms with $\nu^2\partial_\nu^2$. However, because of Eq.~\eqref{eq:gen_kernel_op_ds_o_nu} we in factor only need the kernel expressions for general Doppler weight.

\subsubsection{Inversion formula and commutation}
\label{sec:inverse}
It is clear that two subsequent boost of the same magnitude but in opposite directions yield the initial field. For the aberration kernel this implies the identity 
\bealf{
\label{eq:inversion_K}
\sum_{\ell'} {^d_s}\mathcal{K}_{\ell\ell'}^{m}(\nu, -\beta)\, {^d_s}\,\mathcal{K}_{\ell'\ell''}^{m}(\nu,\beta)&=\delta_{\ell\ell''}.
}
Similarly, the boost operator follows the identity
\bealf{
\label{eq:inversion_B}
\sum_{\ell'} {^d_s}\mathcal{\hat{B}}_{\ell\ell'}^{m}(\nu, -\beta)\, {^d_s}\,\mathcal{\hat{B}}_{\ell'\ell''}^{m}(\nu,\beta)&=\delta_{\ell\ell''}.
}
It is also easy to show that different boost operators commute: 
\bealf{
\label{eq:inversion_B}
{^d_s}\mathcal{\hat{B}}_{\ell\ell'}^{m}(\nu, \beta)\,{^d_s}\mathcal{\hat{B}}_{\ell''\ell'''}^{m}(\nu, \beta)={^d_s}\mathcal{\hat{B}}_{\ell''\ell'''}^{m}(\nu, \beta)\,{^d_s}\mathcal{\hat{B}}_{\ell\ell'}^{m}(\nu, \beta),
}
since the frequency operator\footnote{Since $\partial_{\ln \nu} \nu^k \partial_\nu^k=\nu^k \partial_\nu^k+\nu^{k+1} \partial_\nu^{k+1}$ one has $\nu^{k+1} \partial_\nu^{k+1}=(\partial_{\ln \nu}-1)\nu^k \partial_\nu^k$. Repeatedly apply this identity confirms $\nu^k \partial_\nu^k \equiv (\partial_{\ln \nu}-1)^{k-1} \partial_{\ln \nu}$ for $k>0$.}
${\nu}^k \partial_\nu^k=(\partial_{\ln \nu}-1)^{k-1} \partial_{\ln \nu}$ for $k>0$
naturally commutes with ${\nu}^{k'} \partial^{k'}_\nu$ once written in terms of $\partial_{\ln \nu}$. This can also be directly concluded by looking at Eq.~\eqref{eq:gen_kernel_op_ds_o_nu}, which means that ${^d_s}\mathcal{\hat{B}}_{\ell\ell'}^{m}$ can be written as a polynomial in the operator $\hat{\mathcal{O}}_\nu$.

Finally, we comment that expressing the boost operator in terms of rapidity $\eta=\tanh^{-1}(\beta)$, a quantity that is additive in special relativity, one directly has 
\bealf{
\label{eq:additivity}
\sum_{\ell'} {^d_s}\mathcal{\hat{B}}_{\ell\ell'}^{m}(\nu, \eta_1) \,{^d_s}\mathcal{\hat{B}}_{\ell'\ell''}^{m}(\nu, \eta_2)
&={^d_s}\mathcal{\hat{B}}_{\ell\ell''}^{m}(\nu, \eta_1+\eta_2)
}
and similarly for ${^d_s}\mathcal{K}_{\ell\ell'}^{m}$. This expression provides another proof for the inversion law, Eq.~\eqref{eq:inversion_B}, since ${^d_s}\mathcal{\hat{B}}_{\ell\ell''}^{m}(\nu, 0)=\delta_{\ell \ell''}$. These relations will be very helpful for the computations carried out below.

\subsubsection{Raising and lowering the Doppler weight and others}
\label{sec:Doppler_raising}
In applications one will often need the boost operator for varying Doppler weight. It is thus useful to obtain Doppler-weight raising and lowering relations. Starting from the definition in Eq.~\eqref{eq:gen_kernel_op_ds} and using the corresponding relations for the aberration kernel \citep[see][for details]{Dai2014}, one finds
\bsub
\bealf{
\label{eq:gen_kernel_op_ds_raise}
{_s^d}\mathcal{\hat{B}}_{\ell\ell'}^{m}&=
\gamma\; {_s^{d-1}}\mathcal{\hat{B}}_{\ell\ell'}^{m}
-\gamma \beta \bigg[
{_s}C^m_{\ell'+1}\,{_s^{d-1}}\mathcal{\hat{B}}_{\ell\ell'+1}^{m}
\nonumber \\
&\qquad\qquad 
+\frac{sm}{\ell'(\ell'+1)}\,{_s^{d-1}}\mathcal{\hat{B}}_{\ell\ell'}^{m}
+{_s}C^m_{\ell'}\,{_s^{d-1}}\mathcal{\hat{B}}_{\ell\ell'-1}^{m}
\bigg] 
\\
\label{eq:lower_d_B}
&=
\gamma\; {_s^{d+1}}\mathcal{\hat{B}}_{\ell\ell'}^{m}
+\gamma \beta \bigg[
{_s}C^m_{\ell+1}\,{_s^{d+1}}\mathcal{\hat{B}}_{\ell+1 \ell'}^{m}
\nonumber \\
&\qquad\qquad 
+\frac{sm}{\ell(\ell+1)}\,{_s^{d+1}}\mathcal{\hat{B}}_{\ell\ell'}^{m}
+{_s}C^m_{\ell}\,{_s^{d+1}}\mathcal{\hat{B}}_{\ell-1\ell'}^{m}
\bigg].
}
\esub
with $\ell \,{_s}C_\ell^m=\sqrt{(\ell^2-m^2)(\ell^2-s^2)/(4\ell^2-1)}$.
In a very similar way {\it all} the other raising and lowering relations (i.e., those for $s$, $m$ and $\ell$ for $d=1$) given in \citet{Dai2014} for the aberration kernel directly apply after switching $\beta\rightarrow -\beta$. 

\vspace{-3mm}
\subsubsection{Raising operations for ${_{s}^{d, k}\mathcal{H}_{\ell\ell '}^{m}}$}
\label{sec:Doppler_raising}
Because ${_s^d}\mathcal{\hat{B}}_{\ell\ell'}^{m}$ is a sum of ${_{s}^{d, k}\mathcal{H}_{\ell\ell '}^{m}} \,\nu^k\partial_\nu^k$, where $\nu^k\partial_\nu^k$ is independent of $s, m, \ell$ and $ \ell'$, all the raising relations valid for ${_s^d}\mathcal{\hat{B}}_{\ell\ell'}^{m}$ also apply to ${_{s}^{d,k}\mathcal{H}_{\ell\ell '}^{m}}$.
For ${_{s}^{d, k}\mathcal{H}_{\ell\ell '}^{m}}$, we can furthermore show that
\bealf{
\label{eq:def_H_gen_k_lower}
{_{s}^{d, k}\mathcal{H}_{\ell\ell '}^{m}}
&=
\int  \!{_{-s}}Y^*_{\ell m}(\vghp) 
\left[\left(\frac{\nu}{\nu'}\right)^{-d}\, 
\left(\frac{\nu}{\nu'}-1\right)^{k} \right]  
{_{-s}}Y_{\ell' m}(\vgh) \id\vghp
\nonumber\\
&=
\int  \!{_{-s}}Y^*_{\ell m}(\vghp) 
\left[\left(\frac{\nu}{\nu'}\right)^{-d}\, 
\left(\frac{\nu}{\nu'}-1\right)^{k-1}\,
\left(\frac{\nu}{\nu'}-1\right) \right]  
{_{-s}}Y_{\ell' m}(\vgh) \id\vghp
\nonumber\\[1mm]
&=
{_{s}^{d-1, k-1}\mathcal{H}_{\ell\ell'}^{m}}
-{_{s}^{d, k-1}\mathcal{H}_{\ell\ell'}^{m}}.
}
Together with Eq.~\eqref{eq:lower_d_B}, we then have
\bealf{
\label{eq:def_H_gen_k_lower}
{_{s}^{d, k}\mathcal{H}_{\ell\ell '}^{m}}
&=
(\gamma-1)\; {_{s}^{d, k-1}\mathcal{H}_{\ell\ell '}^{m}}
+\gamma \beta 
\bigg[
{_s}C^m_{\ell+1}\,
{_{s}^{d, k-1}\mathcal{H}_{\ell+1 \ell '}^{m}}
\nonumber \\
&\qquad\qquad 
+\frac{sm}{\ell(\ell+1)}\,
{_{s}^{d, k-1}\mathcal{H}_{\ell \ell '}^{m}}
+{_s}C^m_{\ell}\,
{_{s}^{d, k-1}\mathcal{H}_{\ell-1 \ell '}^{m}}
\bigg].
}
We note that we could not identify a $k$-lowering relation.

Starting with ${_{s}^{d, 0}\mathcal{H}_{\ell\ell '}^{m}}(\beta)\equiv {_s^{d}\mathcal{K}_{\ell\ell'}^{m}(-\beta)}$ and apply Eq.~\eqref{eq:def_H_gen_k_lower}
one can then obtain ${_{s}^{d, k}\mathcal{H}_{\ell\ell '}^{m}}(\beta)$ for all $k>0$ once ${_s^{d}\mathcal{K}_{\ell\ell'}^{m}(-\beta)}$ is known. This procedure reduces the number of operations in comparisons to using approaches based on Eq.~\eqref{eq:def_H_gen} with the corresponding $d$-changing relations and it also minimizes numerical cancellation issues.

\vspace{-3mm}
\section{Differential equation formalism}
As shown by \citet{Dai2014}, the aberration kernel can be efficiently computed using an ordinary differential equation formulation. We now generalize this approach to arbitrary Doppler weight and the computations of the boost operator. 

\subsection{Generalized differential equation for $^{d}_s\mathcal{K}_{\ell\ell'}^{m}$}
\label{sec:K_ODE}
The {\it boost operator} depends directly on the aberration kernel which can be explicitly constructed as \citep{Dai2014}
\bealf{
^{1}_s\mathcal{K}_{\ell\ell'}^{m}(\eta) &=\left<s \ell m \left| \,\expf{i \eta \hat{Y}_z} \,\right| s \ell' m\right>
}
for Doppler weight $d=1$. Here we used the 'bra-ket' notation for the projection integrals over the aberration operator, $\hat{\mathcal{A}}_z(\eta)=\expf{i \eta \hat{Y}_z}$. In this expression, the {\it aberration generator} is given by
\bealf{
\hat{Y}_z&=-i \left[\mu+(\mu^2-1)\partial_\mu\right].
}
Following the arguments of Sect.~III.B in \citet{Dai2014}, it is easy to show that the formalism can be generalized to include varying Doppler weight by introducing the operator
\bealf{
{^d}\hat{Y}_z&=-i \left[\mu\,d+(\mu^2-1)\partial_\mu\right]\equiv \hat{Y}_z-i \mu\,(d-1),
}
which then implies
\bealf{
^{d}_s\mathcal{K}_{\ell\ell'}^{m}(\eta) &=\left<s \ell m \left| \,\expf{i \eta \;{^d}\hat{Y}_z} \,\right| s \ell' m\right>.
}
This provides the opportunity to compute the kernel for any Doppler weight using and ordinary differential equation approach. For Doppler weight $d=1$, one can find \citep{Dai2014}
\bealf{
\partial_\eta \,{^1_s}\mathcal{K}_{\ell\ell'}^{m}(\eta) &={_s}B^m_{\ell'+1}\,{^1_s}\mathcal{K}_{\ell\ell'+1}^{m}(\eta)-{_s}B^m_{\ell'}\,{^1_s}\mathcal{K}_{\ell \ell'-1}^{m}(\eta).
}
with $_s B_\ell^m\equiv \ell \,{_s}C_\ell^m=\sqrt{(\ell^2-m^2)(\ell^2-s^2)/(4\ell^2-1)}$. This differential equation system readily generalizes to
\bealf{
\label{eq:KD_gen}
\partial_\eta \,{^d_s}\mathcal{K}_{\ell\ell'}^{m} &=
\left<s \ell m \left| \, \expf{i \eta \,{^{d}}\hat{Y}_z} \left[i \hat{Y}_z+(d-1)\,\mu\right]\,\right| s \ell' m\right>
\nonumber\\[1mm]
&=
{_s}B^m_{\ell'+1}\,{^d_s}\mathcal{K}_{\ell \ell'+1}^{m}-{_s}B^m_{\ell'}\,{^d_s}\mathcal{K}_{\ell \ell'-1}^{m}
\nonumber\\
&\!\!\!+
(d-1)\left[{_s}C^m_{\ell'+1}\,{^d_s}\mathcal{K}_{\ell \ell'+1}^{m}
+\frac{sm}{\ell'(\ell'+1)}\,{^d_s}\mathcal{K}_{\ell \ell'}^{m}
+{_s}C^m_{\ell'}\,{^d_s}\mathcal{K}_{\ell \ell'-1}^{m}\right]
\nonumber
\\[1mm]
&=
(\ell'+d)\,{_s}C^m_{\ell'+1}\,{^d_s}\mathcal{K}_{\ell \ell'+1}^{m}
-\frac{sm(1-d)}{\ell'(\ell'+1)}\,{^d_s}\mathcal{K}_{\ell \ell'}^{m}
\nonumber \\
&\qquad\qquad\qquad\qquad  
-(\ell'+1-d)\,{_s}C^m_{\ell'}\,{^d_s}\mathcal{K}_{\ell \ell'-1}^{m}
}
for any Doppler weight $d$. Here we used the standard identities
\bealf{
i \hat{Y}_z \,\big| s \ell m\big>&={_s}B^m_{\ell+1}\,\big| s \, \ell+1 \,m\big>-{_s}B^m_{\ell}\,\big| s \,\ell-1 \,m\big>
\nonumber
\\
\nonumber
\mu\,\big| s \ell m\big>&={_s}C^m_{\ell+1}\,\big|  s\, \ell+1 \,m\big>+\frac{sm}{\ell(\ell+1)}\,\big| s\ell m\big>+{_s}C^m_{\ell}\,\big| s \,\ell-1 \,m\big>.
}
Note that $\mu$ and $i \hat{Y}_z$ do not {\it individually} commute but ${^{d}}\hat{Y}_z$ commutes with itself.
Equation~\eqref{eq:KD_gen} can be solved numerically with the initial condition ${^d_s}\mathcal{K}_{\ell \ell'}^{m}(0)=\delta_{\ell\ell'}$ \citep[see][for more details]{Dai2014}, providing a highly efficient and numerically stable procedure that avoids expensive recursion relations \citep[e.g.,][]{Chluba2011ab}. 

\vspace{-3mm}
\subsection{Differential equation for the boost operator}
\label{sec:ODE_for_B}
With Eq.~\eqref{eq:gen_kernel_op_ds}, we can compute ${_s^d}\mathcal{\hat{B}}_{\ell\ell'}^{m}$ once all required ${_{s}^{d, k}\mathcal{H}_{\ell\ell '}^{m}}$ are known. Since with Eq.~\eqref{eq:KD_gen} we have a differential equation for ${^d_s}\mathcal{K}_{\ell\ell'}^{m}$ (i.e., general Doppler weight), and because ${_{s}^{d, 0}\mathcal{H}_{\ell\ell '}^{m}}\equiv {^d_s}\mathcal{K}_{\ell\ell'}^{m}$, we can compute all the required ${_{s}^{d, k}\mathcal{H}_{\ell\ell '}^{m}}$ by applying Eq.~\eqref{eq:def_H_gen_k_lower}. Overall, this already reduces the total computational burden in obtaining the boost operator. We note that to improve stability, in numerical computations it is beneficial to write $\gamma-1=\gamma^2 \beta^2/[\gamma+1]$.

However, for analytical applications, the intermediate step of computing ${_{s}^{d,k}}\mathcal{H}_{\ell \ell'}^{m}$ can be avoided, because one can directly obtain a formal differential equation for ${_s^d}\mathcal{\hat{B}}_{\ell\ell'}^{m}$. 
Since we have 
\bealf{
{_s^d}\mathcal{\hat{B}}_{\ell\ell'}^{m}&\approx 
\left<s \ell m \left| \, 1 - \eta \left[\mu\left(d+\hat{\mathcal{O}}_\nu\right)+(\mu^2-1)\,\partial_\mu \right] \,\right| s \ell' m\right>
}
for infinitesimal rapidity, we can directly write
\bsub
\bealf{
{_s^d}\mathcal{\hat{B}}_{\ell\ell'}^{m}&= 
\left<s \ell m \left| \, \expf{-i \eta \;{^d}\hat{Y}_{z,\nu}} \,\right| s \ell' m\right>
\\
{^d}\hat{Y}_{z,\nu}&=-i \left[\mu\,(d+\hat{\mathcal{O}}_\nu)+(\mu^2-1)\partial_\mu\right] = {^d}\hat{Y}_z-i \mu \, \hat{\mathcal{O}}_\nu,
}
\esub
where ${^d}\hat{Y}_{z,\nu}$ is the {\it boost generator} along the $z$-axis. As anticipated in Sec.~\ref{sec:Alternative_Onu}, this shows that one can simply think of the boost operator in terms of the  aberration generator with $d\rightarrow d+\hat{\mathcal{O}}_\nu$. 
By flipping the sign of $\eta$ (as it is not related to an observer operator), from Eq.~\eqref{eq:KD_gen} we can then immediately write\footnote{Note the overall flip of sign due to $\partial_{-\eta}{^d_s}\mathcal{K}_{\ell \ell'}^{m}(-\eta)=-\partial_{\eta}{^d_s}\mathcal{K}_{\ell \ell'}^{m}(-\eta)$.}
\bealf{
\label{eq:BD_gen}
\partial_\eta \,{^d_s}\mathcal{\hat{B}}_{\ell\ell'}^{m} 
\nonumber
&=
-(\ell'+d+\hat{\mathcal{O}}_\nu)\,{_s}C^m_{\ell'+1}\,{^d_s}\mathcal{\hat{B}}_{\ell \ell'+1}^{m}
+\frac{sm\,(1-d-\hat{\mathcal{O}}_\nu)}{\ell'(\ell'+1)}\,{^d_s}\mathcal{\hat{B}}_{\ell \ell'}^{m}
\\
&\qquad\quad
+(\ell'+1-d-\hat{\mathcal{O}}_\nu)\,{_s}C^m_{\ell'}\,{^d_s}\mathcal{\hat{B}}_{\ell \ell'-1}^{m}.
}
This is a {\it formal} differential equation for the boost operator that can be used to produce term by term in the Taylor series, by repeatedly applying the expression to obtain higher order derivatives with respect to $\eta$, as we illustrate next.

\subsection{Taylor series for the boost operator to second order in $\beta$}
To demonstrate how to apply Eq.~\eqref{eq:BD_gen}, let us consider the Taylor series for ${^d_s}\mathcal{\hat{B}}_{\ell \ell'}^{m}$ to second order in $\beta$. For $|\eta|\ll 1$, we require
\bealf{
\label{eq:BD_gen_Taylor}
{^d_s}\mathcal{\hat{B}}_{\ell\ell'}^{m}(\nu, \eta) 
&\approx {^d_s}\mathcal{\hat{B}}_{\ell\ell'}^{m}\Big|_{\eta=0}+
\partial_\eta {^d_s}\mathcal{\hat{B}}_{\ell\ell'}^{m}\Big|_{\eta=0}\,\eta+
\partial^2_\eta \, {^d_s}\mathcal{\hat{B}}_{\ell\ell'}^{m}\Big|_{\eta=0}\,\frac{\eta^2}{2}.
}
Since $\eta\approx \beta + \mathcal{O}(\beta^3)$ we do not need to worry about the differences between $\eta$ and $\beta$ at second order. Using Eq.~\eqref{eq:BD_gen}, we then have
\bsub
\bealf{
{^d_s}\mathcal{\hat{B}}_{\ell\ell'}^{m}\Big|_{\eta=0}&=\delta_{\ell\ell'}
\\[2mm]
\partial_\eta \,{^d_s}\mathcal{\hat{B}}_{\ell\ell'}^{m}\Big|_{\eta=0} 
&=
-(\ell'+d+\hat{\mathcal{O}}_\nu)\,{_s}C^m_{\ell'+1}\,\delta_{\ell \ell'+1}
\nonumber 
+\frac{sm\,(1-d-\hat{\mathcal{O}}_\nu)}{\ell'(\ell'+1)}\,\delta_{\ell \ell'}
\\
&\qquad\qquad
+(\ell'+1-d-\hat{\mathcal{O}}_\nu)\,{_s}C^m_{\ell'}\,\delta_{\ell \ell'-1}
\\[2mm]
\label{eq:d2Bdeta}
\partial^2_\eta \,{^d_s}\mathcal{\hat{B}}_{\ell\ell'}^{m}\big|_{\eta=0}  
&=(\ell'+d+\mathcal{\hat{O}}_\nu)(\ell'+1+d+\mathcal{\hat{O}}_\nu)\,{_s}C^m_{\ell'+1}\,{_s}C^m_{\ell'+2}\,\delta_{\ell \ell'+2}
\nonumber\\
&\!\!\!\!\!\!\!\!\!\!\!\!\!\!\!\!\!\!\!\!\!\!\!\!
-\frac{2sm}{\ell'(\ell'+2)}
\,(1-d-\mathcal{\hat{O}}_\nu)(\ell'+d+\mathcal{\hat{O}}_\nu)\,{_s}C^m_{\ell'+1}\,\delta_{\ell \ell'+1}
\\ \nonumber
&\!\!\!\!\!\!\!\!\!\!\!\!\!\!\!\!\!\!\!\!
-\Bigg\{
(\ell'+d+\mathcal{\hat{O}}_\nu)
(\ell'+2-d-\mathcal{\hat{O}}_\nu)\,({_s}C^m_{\ell'+1})^2 
-\frac{s^2m^2\,(1-d-\mathcal{\hat{O}}_\nu)^2}{{\ell'}^2(\ell'+1)^2}
\\
&\!\!\!\!\!\!\!\!\!\!\!\!\qquad\quad
+(\ell'+1-d-\mathcal{\hat{O}}_\nu)
(\ell'-1+d+\mathcal{\hat{O}}_\nu)\,({_s}C^m_{\ell'})^2
\Bigg\}\,\delta_{\ell \ell'}
\nonumber \\
&\!\!\!\!\!\!\!\!
+\frac{2sm}{(\ell'-1)(\ell'+1)}
\,(\ell'+1-d-\mathcal{\hat{O}}_\nu)(1-d-\mathcal{\hat{O}}_\nu)\,{_s}C^m_{\ell'}\,\delta_{\ell \ell'-1}
\nonumber \\ \nonumber 
&
+
(\ell'+1-d-\mathcal{\hat{O}}_\nu)(\ell'-d-\mathcal{\hat{O}}_\nu)\,{_s}C^m_{\ell'-1}\,{_s}C^m_{\ell'}\,\delta_{\ell \ell'-2}.
}
\esub
The intermediate steps in obtaining $\partial^2_\eta \,{^d_s}\mathcal{\hat{B}}_{\ell\ell'}^{m}\big|_{\eta=0}$ are spelled out in Appendix~\ref{app:B_coefficies}. 

In the derivations below we will need the relation for $s=0$, which can be simplified to 
\bealf{
\label{eq:BD_gen_Taylor_final}
{^d}\mathcal{\hat{B}}_{\ell\ell'}^{m}(\nu, \beta) 
&= {^d_0}\mathcal{\hat{B}}_{\ell\ell'}^{m}(\nu, \beta) \approx 
\nonumber 
\\
&\!\!\!\!\!\!\!\!\!\!\!\!\!\!\!\!
\left[1-\Bigg\{
\left[
(\ell'+1)^2-{^d}\mathcal{\hat{O}}^2_\nu\right]\,(C^m_{\ell'+1})^2 
+\left[{\ell'}^2-{^d}\mathcal{\hat{O}}^2_\nu\right]\,(C^m_{\ell'})^2 
\Bigg\}\,\frac{\beta^2}{2}\right]\delta_{\ell \ell'}
\nonumber 
\\[1mm]
&\!\!\!\!\!\!\!\!\!\!\!\!\!\!\!\!\!\!\!\!\!
- (\ell'+d+\hat{\mathcal{O}}_\nu)\,C^m_{\ell'+1}\,\beta\,\delta_{\ell \ell'+1}
+(\ell'+1-d-\hat{\mathcal{O}}_\nu)\,C^m_{\ell'}\,\beta\,\delta_{\ell \ell'-1} 
\nonumber 
\\[1mm]
&\!\!\!\!\!\!\!\!\!\!\!\!\!\!\!\!\!\!\!\!\!
+C^m_{\ell'+1} C^m_{\ell'+2}
\left[(\ell'+d)(\ell'+1+d)+2(\ell'+2+d)\,\mathcal{\hat{O}_\nu}+\oDnu\right] 
\frac{\beta^2}{2}
\delta_{\ell \ell'+2}
\nonumber\\
&\!\!\!\!\!\!\!\!\!\!\!\!\!\!\!\!
+C^m_{\ell'-1}C^m_{\ell'} 
\left[(\ell'-d)(\ell'+1-d)-2(\ell'-1-d)\,\mathcal{\hat{O}_\nu}+\oDnu\right] 
\frac{\beta^2}{2}
\delta_{\ell \ell'-2}
\nonumber 
\\[2mm]
{^d}\mathcal{\hat{O}}^2_\nu
&=(d-1+\mathcal{\hat{O}_\nu})^2=(1-d)^2+2(d-1)\mathcal{\hat{O}_\nu}+\mathcal{\hat{O}}^2_\nu
\nonumber \\
&=(1-d)^2-(2d-3)\nu\partial_\nu+\nu^2\partial^2_\nu
\nonumber \\
&=(1-d)^2+(2d+1)\,\mathcal{\hat{O}}_\nu+\oDnu
}
up to second order in $\beta$. Here $C^m_{\ell'}={_0}C^m_{\ell'}=\sqrt{(\ell^2-m^2)/(4\ell^2-1)}$ and $\oDnu=\mathcal{\hat{O}}_\nu^2-3\mathcal{\hat{O}}_\nu=\nu^{-2}\partial_\nu \nu^4 \partial_\nu$ is the energy diffusion operator that appears in the Kompaneets equation \citep{chluba_spectro-spatial_2023-I}.

Adding higher order terms is straightforward, but we shall stop here. One can clearly see that coupling with $\Delta\ell=\ell'-\ell$ propagates at order $\beta^{|\Delta \ell |}$ \citep{Chluba2011ab}. Thus, at a fixed order in $\beta$ only a finite range in $\ell$ is affected. 
We also confirmed that the result given in Eq.~\eqref{eq:BD_gen_Taylor} has all the properties summarized in Sect.~\ref{sec:properties_boost_op}.
For the $\ell$-switching symmetry, Eq.~\eqref{eq:gen_kernel_op_ds_o_nu_switch}, one obtains ${_s^d}\mathcal{\hat{B}}_{\ell'\ell}^{m}(\nu, \beta)$ from ${_s^d}\mathcal{\hat{B}}_{\ell\ell'}^{m}(\nu, \beta)$ by replacing $d\rightarrow 2-d-2\oOnu$ and $\beta\rightarrow -\beta$.

\section{Applications of the boost operator}
In this section, we apply the boost operator approach to a few illustrative examples. We first briefly recap how the CMB anisotropies for $\ell\leq 2$ transform and how motion-induced effects propagate. 
We then move on to the kinetic equation of Compton scattering, for which we start in the Thomson limit and then generalize to include terms to first order in the electron temperature.

We will work with photon occupation number (i.e., $s=d=0$). From Eq.~\eqref{eq:BD_gen_Taylor_final}, with $B^m_{\ell}={_0}B^m_{\ell}=\ell C^m_{\ell}$ we then have
\bealf{
\label{eq:BD_gen_Taylor_final_00}
\mathcal{\hat{B}}_{\ell\ell'}^{m}(\nu, \beta) 
&= {^0_0}\mathcal{\hat{B}}_{\ell\ell'}^{m}(\nu, \beta) \approx 
\nonumber 
\\
&\!\!\!\!\!\!\!\!\!\!\!\!\!\!\!\!
\left[1-\Bigg\{
(B^m_{\ell'+1})^2+(B^m_{\ell'})^2
-{^0}\mathcal{\hat{O}}^2_\nu\,\left[(C^m_{\ell'+1})^2+(C^m_{\ell'})^2\right] 
\Bigg\}\,\frac{\beta^2}{2}\right]\delta_{\ell \ell'}
\nonumber 
\\[1mm]
&\!\!\!\!\!\!\!\!\!
- (\ell'+\hat{\mathcal{O}}_\nu)\,C^m_{\ell'+1}\,\beta\,\delta_{\ell \ell'+1}
+(\ell'+1-\hat{\mathcal{O}}_\nu)\,C^m_{\ell'}\,\beta\,\delta_{\ell \ell'-1} 
\nonumber 
\\[1mm]
&\!\!\!\!\!\!\!\!\!
+C^m_{\ell'+1} C^m_{\ell'+2}
\left[\ell'(\ell'+1)+2(\ell'+2)\,\mathcal{\hat{O}_\nu}+\oDnu\right] 
\frac{\beta^2}{2}
\delta_{\ell \ell'+2}
\nonumber\\
&\!\!\!\!
+C^m_{\ell'-1}C^m_{\ell'} 
\left[\ell'(\ell'+1)-2(\ell'-1)\,\mathcal{\hat{O}_\nu}+\oDnu\right] 
\frac{\beta^2}{2}
\delta_{\ell \ell'-2}
\nonumber 
\\[2mm]
{^0}\mathcal{\hat{O}}^2_\nu
&=1-2\mathcal{\hat{O}_\nu}+\mathcal{\hat{O}}^2_\nu
=1+3\nu\partial_\nu+\nu^2\partial^2_\nu
=1+\mathcal{\hat{O}}_\nu+\oDnu
}
up to second order in $\beta$. We note that one can alternatively work with photon intensities, $I_\nu\propto \nu^3 n(\nu)$. In this case the expressions for $s=0$ and $d=3$ have to be used. However, the frequency-integrated intensity is most closely related to the observations of the sky, meaning that one has to consider cases for $s=0$ and $d=4$. The latter can also be obtained from the transformed photon occupation number. 

\subsection{CMB monopole, dipole and quadrupole and their spectra}
\label{sec:dipole_quadrupole}
Since the Solar system is moving with respect to the CMB rest frame, we can use the boost operator approach to explicitly compute the spectra of the lowest multipoles, here up to $\ell=2$. 
Given that at second order in $\beta$ multipoles with $\Delta \ell \leq 2$ mix, an observer moving with respect to the CMB will find that the CMB monopole through quadrupole ($\ell\leq 2$) are generally described by a mixture of the spectra and their derivatives from multipoles $\ell\leq 4$. 
To model the observational process, one will also have to compute the frequency-integrated quantities and distinguish between foregrounds that are comoving with the observer, as these will not have to be transformed \citep{Dai2014}, complicating the mixing process significantly.

To prepare the discussion, we shall start by giving the explicit expressions for the transformed CMB to second order in $\beta$. With the boost operator for $s=d=0$ it is easy to obtain the corresponding photon occupation number from Eq.~\eqref{eq:n_mov_final} and \eqref{eq:BD_gen_Taylor_final} for an observer ($\beta \rightarrow -\beta$) in the moving frame:
\bealf{
\label{eq:motion_CMB}
n'_{00}(\nu)
&\approx \left[1+\frac{\beta^2}{6}\left(\oOnu+\oDnu \right)\right] n_{00}(\nu)
-\frac{\beta}{\sqrt{3}} \left[2-\oOnu\right] n_{10}(\nu)
\nonumber \\
&\qquad\qquad+\frac{\beta^2}{3\sqrt{5}}\left[6-2\oOnu+\oDnu \right] n_{20}(\nu)
\\[1mm]
n'_{10}(\nu)
&\approx \frac{\beta}{\sqrt{3}} \oOnu n_{00}(\nu)
+\left[1-\frac{\beta^2}{10}\left(4-3\oOnu-3\oDnu] \right)\right]  n_{10}(\nu)
\nonumber \\
&\!\!\!\!
-\frac{2\beta}{\sqrt{15}}\left[3-\oOnu\right] n_{20}(\nu)
+\frac{3\beta^2}{5\sqrt{21}}\left[12-4\oOnu+\oDnu \right] n_{30}(\nu)
\nonumber \\[1mm]
n'_{1\pm1}(\nu)
&\approx
\left[1-\frac{\beta^2}{10}\left(3-\oOnu-\oDnu\right)\right]  n_{1\pm1}(\nu)
\nonumber \\
&\!\!\!\!
-\frac{\beta}{\sqrt{5}}\left[3-\oOnu\right] n_{2\pm1}(\nu)
+\frac{2\beta^2}{5\sqrt{14}}\left[12-4\oOnu+\oDnu \right] n_{3\pm1}(\nu)
\nonumber 
\\[1mm]
\nonumber
n'_{20}(\nu)
&\approx 
\frac{\beta^2}{3\sqrt{5}} \left[4\oOnu+\oDnu\right] n_{00}(\nu)
+\frac{2\beta}{\sqrt{15}} \left[1+\oOnu\right] n_{10}(\nu)
\nonumber\\ \nonumber
&\!\!\!\!+\left[1-\left(\frac{10}{7}-\frac{11}{42}\oOnu-\frac{11}{42}\oDnu\right)\beta^2\right] n_{20}(\nu)
\\ \nonumber
&
-\frac{3\beta}{\sqrt{35}} \left[4-\oOnu\right] n_{30}(\nu)
+\frac{2\beta^2}{7\sqrt{5}} \left[20-6\oOnu+\oDnu\right] n_{40}(\nu).
\\ \nonumber
n'_{2\pm1}(\nu)
&\approx 
\frac{\beta}{\sqrt{5}} \left[1+\oOnu\right] n_{1\pm1}(\nu)
\nonumber\\ \nonumber
&\!\!\!\!+\left[1-\left(\frac{17}{14}-\frac{3}{14}\oOnu-\frac{3}{14}\oDnu\right)\beta^2\right] n_{2\pm1}(\nu)
\\ \nonumber
&
-\frac{4\beta}{\sqrt{70}} \left[4-\oOnu \right] n_{3\pm1}(\nu)
+\frac{2\beta^2}{7\sqrt{6}} \left[20-6\oOnu+\oDnu\right] n_{4\pm1}(\nu).
\\ \nonumber
n'_{2\pm2}(\nu)
&\approx 
\left[1-\left(\frac{4}{7}-\frac{1}{14}\oOnu-\frac{1}{14}\oDnu\right)\beta^2\right]n_{2\pm2}(\nu)
\\ \nonumber
&
-\frac{\beta}{\sqrt{7}} \left[4-\oOnu\right] n_{3\pm2}(\nu)
+\frac{\beta^2}{7\sqrt{3}} \left[20-6\oOnu+\oDnu\right] n_{4\pm2}(\nu).
}
We can clearly see how the spectra at a given $\ell$ receive contributions from $\ell\pm1$ at first and $\ell\pm2$ at second order in $\beta$. This also means that measurements of the monopole through quadrupole in the moving frame are not independent of the octupole and hexadecupole  in the resting frame, giving rise to truncation issues. To appreciate the relevance of the various terms, we now consider a few examples. 

\subsubsection{Kinematic dipole and quadrupole}
Assuming that in the CMB rest frame the spectrum is given by an average monopole only (i.e., $n_{\ell m}(\nu)=n_{00}(\nu)\delta_{\ell 0}\delta_{m0}$), from Eq.~\eqref{eq:motion_CMB} we recover the well-known kinematic dipole and quadrupole,
\bealf{
\label{eq:transformed_sky}
n'_{0}(\nu)
&\approx \left[1+\frac{\beta^2}{6}\left(\oOnu+\oDnu\right)\right] \bar{n}_{0}(\nu)
=\bar{n}_{0}(\nu)+\frac{\beta^2}{6}[G(x)+Y(x)]
\nonumber
\\
n'_{10}(\nu')&\approx \frac{\beta}{\sqrt{3}} \,\oOnu n_{00}(\nu)=\sqrt{\frac{4\pi}{3}}\,\beta\,G(x)
\\ \nonumber
n'_{20}(\nu')&\approx 
\frac{\beta^2}{3\sqrt{5}} \left[4\oOnu+\oDnu\right] n_{00}(\nu)
=\sqrt{\frac{4\pi}{5}}\,\frac{\beta^2}{3}\,[4G(x)+Y(x)],
}
with $x=h\nu/kT_0$ and $ \bar{n}_{0}(\nu)=1/(\expf{x}-1)$. We furthermore used $n_{00}=\sqrt{4\pi}\,\bar{n}_{0}$ and $n'_{0}=n'_{00}/\sqrt{4\pi}$ to simplify the notation. The CMB monopole temperature is $T_0=2.725\,\Kel$ \citep{Fixsen1996, Fixsen2009} and the function $G(x)=-\nu\partial_\nu  \bar{n}_0 =x\expf{x}/(\expf{x}-1)^2$ describes the spectrum of a temperature fluctuation.

As Eq.~\eqref{eq:transformed_sky} demonstrates, 
a CMB cosmologist receives blackbody photons with a slightly higher temperature in the forward direction, finding a temperature variation $\Delta T_1/T_0= \beta \simeq \pot{1.241}{-3}$ \citep{Smoot1977, Fixsen1996}. 
In addition, we see that the effective temperature of the CMB monopole is increased by $\Delta T/T_0\approx \beta^2/6$. 
The monopole and quadrupole spectra furthermore receive a small $y$-type distortion with a $y$-parameter $y=\beta^2/6$ and distortion spectrum $Y(x)=\oDnu \bar{n}_0=G(x)\left[x\coth(x/2)-4\right]$.

The origin of all these terms is due to boosting effects and the superposition of blackbody spectra when performing the spherical harmonic expansion of the field \citep[e.g.,][]{Kamionkowski2003, Chluba2004,Chluba2011ab, Sunyaev2013dipole}. To see this, let us write the transformed CMB
\bealf{
\label{eq:CMB_transformed}
n'(\nu', \vghp)&\equiv n[\nu(\nu',\vghp), \vgh(\vghp)]=\frac{1}{\expf{ \gamma x (1-\beta \mu')}-1},
}
which is explicitly given by a blackbody field with varying temperature $T'(\vghp)=T_0/[\gamma (1-\beta \mu')]\approx T_0[1-\beta^2/2+\beta \mu'+\beta^2 \mu'^2]$. 
Taking the average of this temperature yields $T'_0\approx T_0(1-\beta^2/6)$ with one contribution from the Lorentz factor ($\Delta T^{\gamma}/T_0=-\beta^2/2$) and a second from Doppler-modulation ($\Delta T^{\rm D}/T_0=\beta^2/3$) \citep[e.g.,][]{Chluba2011ab}; however, the superposition of blackbody spectra adds $\Delta T^{\rm sup}/T_0=\beta^2/3$ \citep{Chluba2004}, yielding a total change of the effective CMB blackbody temperature by $\Delta T/T_0\approx \beta^2/6$.

By measuring the dipole amplitude we can therefore predict the $y$-distortion of the monopole and quadrupole spectra. 
Even if one can in principle extract the related $y$-parameter using CMB measurements \citep[e.g.,][]{Kamionkowski2003, Chluba2004}, 
we note that this distortion can be entirely avoided by first converting the CMB fluxes in each pixel into the thermodynamic temperature and then performing the spherical harmonic expansion on the temperature sky \citep[e.g.,][]{Chluba2017foregrounds}. As such, the related monopole and quadrupole distortions do not add any {\it new} information to the system aside from indicating that we have not worked in temperature space, a correction that is, however, relevant to precise CMB analyses \citep{Sullivan2021}. 

We also note that in \citet{Chluba2016} the Doppler-modulation contribution $\Delta T^{\rm D}/T_0=\beta^2/3$ was erroneously omitted, yielding an overall sign change of the total contribution. However, since the superposition of blackbody contribution can be avoided, a measurement of the average {\it thermodynamic} temperature would indeed yield $\Delta T/T_0\approx -\beta^2/6$ as the $\Lambda$CDM contribution from the motion-induced CMB dipole, as correctly stated there.

\subsubsection{Searching for the primordial CMB dipole}
Past considerations \citep{Colin2019} have highlighted how important it could be to obtain independent limits on the amplitude of the primordial dipole [e.g., produced from large-scale isocurvature perturbations \citep[see][and references therein]{Zibin2008}]. Further motivation is provided due to the presence of large-angle anomalies \citep[e.g.,][]{Copi2006, Copi2010, Dai2013}, which could be connected to the close alignment of the two dipoles. 

The question then is, can one use the information in the lowest multipoles to extract the primordial dipole when taking all effects into account? Let us first write the CMB with a primordial temperature dipole consistent to second order in $\Delta_T=\Delta T/T$:
\bealf{
n(\nu, \vgh)\approx \frac{1}{\expf{x}-1}+\left[\Delta_T(\vgh)+\Delta^2_T(\vgh)\right]\,G(x)+\frac{\Delta^2_T(\vgh)}{2}\,Y(x),
}
with $\Delta_T(\vgh)=\sum_{m=-1}^{1}\,a_{1m}\,Y_{1m}(\vgh)$ and $a_{1-m} = (-1)^m  a^*_{1m}$. Performing a spherical harmonic transformation of this expression then gives
\bsub
\label{eq:CMB_with_prim_dipole}
\bealf{
n_0&=\frac{n_{00}}{\sqrt{4\pi}}=\bar{n}_{0}+\frac{\beta_{\rm pr}^2}{3}\,\left[G(x)+\frac{1}{2}\,Y(x)\right],
\\
n_{1m}&= a_{1m}\,G(x),
\\
n_{20}&= 
\frac{|a_{10}|^2-|a_{11}|^2}{\sqrt{5\pi}}\,\left[G(x)+\frac{1}{2}\,Y(x)\right],
\\
n_{2\pm1}&= 
\sqrt{\frac{3}{5\pi}}\,a_{10}\,a_{1\pm 1}\,\left[G(x)+\frac{1}{2}\,Y(x)\right],
\\
n_{2\pm2}&= 
\sqrt{\frac{3}{10\pi}}\,a_{1\pm 1}^2\,\left[G(x)+\frac{1}{2}\,Y(x)\right],
}
\esub
with the effective primordial dipole velocity amplitude $$\beta^2_{\rm pr}=\frac{3}{4\pi}\sum_{m=-1}^{1}\,|a_{1m}|^2.$$
We can see that at second order in $\Delta_T$ the spectra of the monopole and quadrupole (here without any primordial quadrupole) are distorted due to the superposition of blackbodies as before. 

By inserting the expressions in Eq.~\eqref{eq:CMB_with_prim_dipole} into those of Eq.~\eqref{eq:motion_CMB} and neglecting third order terms, we then find 
\bealf{
\label{eq:transformed_sky_both}
n'_{0}(\nu)
&\approx
\bar{n}_{0}(\nu)
+\left[\frac{\beta^2}{6}+\frac{\beta_{\rm pr}^2}{3}+\frac{\beta\,a_{10}}{2\sqrt{3\pi}}\right]G(x)
\nonumber \\
&\qquad\qquad\qquad +\left[\frac{\beta^2}{3}+\frac{\beta_{\rm pr}^2}{3}+\frac{\beta\,a_{10}}{\sqrt{3\pi}}\right]\frac{Y(x)}{2}
\nonumber
\\[1mm] \nonumber
n'_{1m}(\nu)
&\approx \sqrt{\frac{4\pi}{3}}\left[\,\beta \,\delta_{m0}+ \sqrt{\frac{3}{4\pi}}\,a_{1 m} \right]G(x)\equiv \int Y^*_{1 m}(\vgh)\, \boldsymbol{\beta}_{\rm t}\cdot\vgh \id\vgh
\\[1mm] \nonumber
n'_{20}(\nu)
&\approx\sqrt{\frac{4\pi}{5}} \left[\frac{4\beta^2}{3}+\frac{4\beta\,a_{10}}{\sqrt{3\pi}}+\frac{|a_{10}|^2-|a_{11}|^2}{2\pi}\right]\,G(x)
\nonumber \\
&\qquad +\sqrt{\frac{4\pi}{5}} \left[\frac{2\beta^2}{3}+\frac{2\beta\,a_{10}}{\sqrt{3\pi}}+\frac{|a_{10}|^2-|a_{11}|^2}{2\pi}\right]\frac{Y(x)}{2}
\\[1mm] \nonumber
n'_{2\pm1}(\nu)
&\approx\sqrt{\frac{3\pi}{5}} \left[\frac{4\beta\,a_{1\pm1}}{\sqrt{3\pi}}+\frac{a_{10}\,a_{1\pm1}}{\pi}\right] G(x)
\nonumber \\ \nonumber
&\qquad +\sqrt{\frac{3\pi}{5}} \left[\frac{2\beta\,a_{1\pm1}}{\sqrt{3\pi}}+\frac{a_{10}\,a_{1\pm1}}{\pi}\right]\frac{Y(x)}{2}
\\[1mm] \nonumber
n'_{2\pm2}(\nu)
&\approx\sqrt{\frac{3}{10\pi}}\,a_{1\pm 1}^2\, \left[G(x)
+\frac{Y(x)}{2}\right],
\nonumber
}
where we used $\oOnu G= \oOnu^2 \bar{n}_0=3 G + Y$. As expected, the coefficients of the dipole are simply given by the sum of the two dipoles, meaning that we can (only) constrain the direction and amplitude of the {\it net} dipole vector $\boldsymbol{\beta}_{\rm t}=\boldsymbol{\beta}+\boldsymbol{\beta}_{\rm pr}$ by measuring $n'_{1m}(\nu)$. 

How are the transformed coefficients of the monopole and quadrupole related to this?
Using the dipole, we can write
\bealf{
\beta_{\rm t}^2 &=\int (\boldsymbol{\beta}_{\rm t}\cdot\vgh)^2 \frac{\id \vgh}{4\pi} \equiv \left[\beta+ \sqrt{\frac{3}{4\pi}}\,a_{1 0} \right]^2+\frac{3}{4\pi}\,|a_{1 1}|^2+\frac{3}{4\pi}\,|a_{1 -1}|^2
\nonumber\\
&=\beta^2+\beta^2_{\rm pr}+\sqrt{\frac{3}{\pi}}\,\beta\,a_{1 0}\equiv 3\left[\frac{\beta^2}{3}+\frac{\beta^2_{\rm pr}}{3}+\frac{\beta\,a_{1 0}}{\sqrt{3\pi}}\,\right],
}
where in the last step we cast the terms into a form that shows that the $y$-distortion parameter of the monopole in Eq.~\eqref{eq:transformed_sky_both} is identical to $y_{\rm t}=\beta_{\rm t}^2/6$. We can therefore write the monopole spectrum as
\bealf{
\label{eq:transformed_sky_both_monopole}
n'_{0}(\nu)
&\approx
\bar{n}_{0}(\nu)
+\left[\frac{\beta_{\rm t}^2}{6}+\frac{\beta_{\rm pr}^2}{6}\right]G(x)
+\frac{\beta_{\rm t}^2}{6} Y(x),
}
which shows that the distortion of the monopole is again just from the (avoidable) superposition of blackbody spectra with no difference with regards to the origin of the dipole. This also implies that by measuring the monopole $y$-distortion amplitude (even in this idealistic setup) we cannot gain any new insights. 

The average CMB monopole temperature furthermore receives an additional contribution $\Delta T/T_0\approx \beta_{\rm pr}^2/6$, which arises as follows: If the effective dipole was purely motion-induced, we would expect $\Delta T/T_0\approx \beta_{\rm t}^2/6$ as in the previous section. However, because $\boldsymbol{\beta}_{\rm pr}$ does not add to the motion, we have to {\it subtract} the purely motion-induced effect, $\Delta T/T_0\approx -\beta_{\rm pr}^2/6$, giving a net positive correction (see Appendix~\ref{app:dipole_transform} for details). Even if this identifies a difference in how motion and intrinsic dipole affect the CMB monopole spectrum, the effect is degenerate with the unknown value of the CMB temperature, $T_0$, implying that we cannot use a combination of monopole and dipole measurements alone to extract information about $\boldsymbol{\beta}_{\rm pr}$.

Can we get to the primordial dipole by including information from the quadrupole? The short answer is {\it no}. This point is apparent when counting degrees of freedom. Ignoring the primordial CMB quadruple and higher order multipoles [as we have in Eq.~\eqref{eq:transformed_sky_both_monopole}], we are dealing with 7 unknowns (1 monopole, 3 primordial dipole and 3 from our motion). 
Using the information from monopole, dipole and quadrupole we have 9 measured parameters, suggesting we should be able to solve the system (even without considering distortion information). However, once we add the primordial CMB temperature quadrupole, we have 12 unknowns, again leaving the system underdetermined when only considering the temperature terms $\propto G(x)$. 

What about using the distortion parts? It turns out that {\it all} the quadrupole distortion terms in Eq.~\eqref{eq:transformed_sky_both} are exactly the spherical harmonic coefficients of  $(\boldsymbol{\beta}_{\rm t}\cdot\vgh)^2/2$ [see Appendix~\ref{app:dipole_transform} and Eq.~\eqref{eq:Quadrupole_general} for details], which we have already constrained more easily by using the CMB dipole information. We therefore do not gain anything new by considering the distortion information \citep[see also discussion in][]{Kamionkowski2003, Sullivan2021}. This statement also does not change when higher order effects are accounted for as no {\it independent} information can be gained.

The degeneracy can be overcome by measuring the breaking of statistical isotropy due to the aberration effect at high multipoles \citep{Burles2006, Amendola2010, Chluba2011ab}. 
The new piece of information put into this is the assumption of statistical isotropy of the primordial CMB anisotropies, in essence defining the 'statistical isotropy frame' as a new gauge.
However, the achievable precision using this method is limited \citep{Planck2013abber} and in addition explicitly relies on the assumption that the CMB is intrinsically statistically isotropic, even if well-justified. 
For the sake of argument, if $\boldsymbol{\beta}_{\rm pr}\simeq -\boldsymbol{\beta}$, we could have the interesting situation that $|\boldsymbol{\beta}_{\rm pr}+\boldsymbol{\beta}|$ is much smaller than the individual contributions. In this case, the motion-induced quadrupole term could in principle receive a contribution that exceeds $2\beta^2 \simeq \pot{2}{-6}$, thus causing an apparent alignment of the observer frame dipole and quadrupole axes, possibly relating to one of the low-$\ell$ CMB anomalies \citep{DiValentino2025Tensions}. From observations of the CMB aberration effect, which provided an independent $5\sigma$ measurement of the motion-induced CMB dipole, we can exclude this possibility \citep{Planck2013abber}, with a current constraint on the primordial dipole, $\beta_{\rm pr}\lesssim 3.7\,{\rm mK}/2.725\,{\rm K} \simeq \pot{1.4}{-3}$ (95\% c.l.), that is similar to $\beta$ \citep{Ferreira2021primDipole}.

In real-world scenarios, CMB observations are obscured by the presence of extra-galactic and galactic foregrounds. These contaminate the observed dipole spectrum at order $\beta$ which allows the corresponding monopole and quadrupole spectra to leak into the dipole \citep{Danese1981dipole, Balashev2015, Trombetti2021}. Similarly, quadrupole and octupole spectra mix by aberration effects, further complicating matters. Differences between (co-moving) galactic and extra-galactic foregrounds and their transformation into our frame also have to be carefully considered, preventing us from 'de-boosting' the observed sky in a simple way \citep{Dai2014, Coulton2022boostSignals}. This certainly complicates the modeling of the CMB sky in detailed CMB analyses, as can also be appreciated from the number of terms in Eq.~\eqref{eq:motion_CMB}.

\subsection{Kinematic corrections to Thomson scattering}
\label{sec:kin_corrs_Thomson}
Another simple application of the boost operator approach is in the derivation of the kinematic corrections to the scattering of photons by moving electrons in the Thomson limit. The collision term for resting electrons is given by:
\bealf{
\label{eq:GFK_n_CS_Thomson}
\left.\frac{\id  n}{ \id \tau}\right|_{\rm T}
&=n_0+\frac{1}{10}\,n_2 - n.
}
where $\tau = \int \sigT \Ne c \id t$ denotes the Thomson optical depth. We also defined $n_\ell(t, \nu, \vgh)=\sum_{m} n_{\ell m}(t, \nu) Y_{\ell m}(\vgh)$ using the spherical harmonic coefficients of the photon occupation number, $n_{\ell m}(t, \nu)$.
These terms all have $s=d=0$ and neglect any energy transfer or Klein-Nishina corrections. Inserting the transformed photon field, we have 
\bealf{
\label{eq:GFK_n_CS_Thomson_prime}
\left.\frac{\id  n'(\nu',\vghp)}{ \id \tau'}\right|_{\rm T}
&=n'_0+\frac{1}{10}\,n'_2 - n'
\nonumber\\
&=\sum_{\ell'} Y_{00}(\vghp)\,{^0}\mathcal{\hat{B}}^0_{0\ell'}(\nu', \beta) \,n_{\ell' 0}(\nu')
\nonumber\\
&\qquad + \sum_{m, \ell'} \frac{Y_{2m}(\vghp)}{10}\,{^0}\mathcal{\hat{B}}^0_{2\ell'}(\nu', \beta) \,n_{\ell' m}(\nu')
\\
\nonumber
&\qquad\qquad - \sum_{\ell, m, \ell'}
Y_{\ell m}(\vghp)\,{^0}\mathcal{\hat{B}}^0_{\ell \ell'}(\nu', \beta) \,n_{\ell' m}(\nu')
}
inside the moving frame. For convenience, we dropped the subscript $s=0$.
To obtain the expression in the lab frame, we perform a boost along the {\it negative} $z$-axis, noting that:
\bealf{
\left.\frac{\id  n(\nu,\vgh)}{ \id \tau}\right|_{\rm T}
&=(1-\beta \mu)\left.\frac{\id  n'\left[\nu'(\nu,\vgh),\vghp(\vgh)\right]}{ \id \tau'}\right|_{\rm T},
}
where\footnote{For a positive boost we have $\nu=\gamma \nu'(1-\beta\mu')$ and thus $\nu'=\gamma \nu(1+\beta\mu)$ and with reversed direction finally $\nu'=\gamma \nu(1-\beta\mu)$.} $\nu'=\gamma \nu (1-\beta\mu)$ and $\mu'=(\mu-\beta)/(1-\beta\mu)$. The overall factor $\nu'/[\gamma \nu]=(1-\beta\mu)$ arises from the transformation of $\id \tau'$, which depends the transformation of the time-interval $\id t'$ and the electron number density $\Ne'=\Ne/\gamma$ \citep[see Appendix~C1 of][for more details]{Chluba2012}. In kinetic theory, this is related to the M{\o}ller relative speed, which essentially {\it lowers} the starting Doppler weight of the previous terms to $d=-1$.

We can then perform a spherical harmonic expansion to obtain the relevant coefficients
\bealf{
\left.\frac{\id  n_{\ell m}(\nu,\vgh)}{ \id \tau}\right|_{\rm T}
&=\int Y^*_{\ell m}(\vgh)(1-\beta \mu)\left.\frac{\id  n'\left[\nu'(\nu,\vgh),\vghp(\vgh)\right]}{ \id \tau'}\right|_{\rm T}\id \vgh
\nonumber 
\\
\nonumber
&=\frac{1}{\gamma} \int Y^*_{\ell m}(\vgh)\left(\frac{\nu'}{\nu}\right)\left.\frac{\id  n'\left[\nu'(\nu,\vgh),\vghp(\vgh)\right]}{ \id \tau'}\right|_{\rm T}\id \vgh.
}
When performing a Taylor series of the $n_{\ell m}(\nu')$ in Eq.~\eqref{eq:GFK_n_CS_Thomson_prime} [this time around $\nu$], we now encounter integrals of the form
\bealf{
{_{k}^{-1}}\mathcal{H}_{\ell\ell'}^{m}(-\beta)
&=\int Y^*_{\ell m}(\vgh)\left(\frac{\nu'}{\nu}\right) \left[\frac{\nu'-\nu}{\nu}\right]^k
Y_{\ell' m}(\vghp)
\id \vgh
\nonumber\\
&= \sum_{t=0}^k \; (-1)^{k-t}\,\binom {\,k\,} {\,t\,}\; {^{-(t+1)}}\mathcal{K}_{\ell\ell'}^{m}(\beta),
\label{eq:def_A}
}
and hence need to use ${^{-1}}\mathcal{\hat{B}}^m_{\ell\ell'}(\nu, -\beta)$. We then have the exact equation
\bealf{
\label{eq:dn_transformed}
\left.\frac{\id  n(\nu,\vgh)}{ \id \tau}\right|_{\rm T}
&=\frac{1}{\gamma}\sum_{\ell,\ell'} Y_{\ell0}(\vgh)\,{^{-1}}\mathcal{\hat{B}}^0_{\ell 0}(\nu, -\beta)\,{^0}\mathcal{\hat{B}}^0_{0\ell'}(\nu, \beta) \,n_{\ell' 0}(\nu)
\nonumber\\
& + \frac{1}{\gamma}\sum_{\ell, m, \ell'} \frac{Y_{\ell m}(\vgh)}{10}
\,{^{-1}}\mathcal{\hat{B}}^m_{\ell 2}(\nu, -\beta)\,{^0}\mathcal{\hat{B}}^m_{2\ell'}(\nu, \beta) \,n_{\ell' m}(\nu)
\\
\nonumber
& - \frac{1}{\gamma}\sum_{\ell, m, \ell', \ell''}Y_{\ell m}(\vgh)
\,{^{-1}}\mathcal{\hat{B}}^m_{\ell \ell'}(\nu, -\beta)\,{^0}\mathcal{\hat{B}}^m_{\ell' \ell''}(\nu, \beta) \,n_{\ell'' m}(\nu)
}
inside the lab frame, where all sums over $m$ are determined by the minimal value of $\ell$ appearing. Formally, this is the final result. However, with Eq.~\eqref{eq:lower_d_B} we can furthermore relate ${^{-1}}\mathcal{\hat{B}}^m_{\ell \ell'}(\nu, -\beta)$ to ${^{0}}\mathcal{\hat{B}}^m_{\ell \ell'}(\nu, -\beta)$.
Introducing ${^{-1}}\mathcal{\hat{T}}^m_{\ell \ell'}(\nu, -\beta)={^{-1}}\mathcal{\hat{B}}^m_{\ell \ell'}(\nu, -\beta)/\gamma$, we find
\bealf{
\label{eq:B_lowering}
{^{-1}}\mathcal{\hat{T}}^m_{\ell \ell'}
&=
{^0}\mathcal{\hat{B}}_{\ell\ell'}^{m}
-\beta \bigg[
C^m_{\ell+1}\,{^0}\mathcal{\hat{B}}_{\ell+1 \ell'}^{m}
+C^m_{\ell}\,{^0}\mathcal{\hat{B}}_{\ell-1\ell'}^{m}
\bigg],
}
where we dropped the arguments for clarity. Using Eq.~\eqref{eq:inversion_B}, thus simplifies the last term in Eq.~\eqref{eq:dn_transformed} to
\bealf{
\nonumber
\sum_{\ell'} 
{^{-1}}\mathcal{\hat{T}}^m_{\ell \ell'}\,{^0}\mathcal{\hat{B}}^m_{\ell' \ell''}
&=\delta_{\ell\ell''}-\beta
\left[C^m_{\ell+1}\,\delta_{\ell+1, \ell''}+C^m_{\ell} \,\,\delta_{\ell-1, \ell''}\right].
}
Since now only the first few terms in Eq.~\eqref{eq:dn_transformed} require evaluations of the boost operator, this significantly reduces the computational burden. 
Using the relations in Eq.~\eqref{eq:BD_gen_Taylor_final}, we can give the explicit expressions for ${^0}\mathcal{\hat{B}}^m_{0\ell}(\nu, \beta)$, ${^0}\mathcal{\hat{B}}^m_{2\ell}(\nu, \beta)$, ${^{-1}}\mathcal{\hat{T}}^m_{\ell0}(\nu, -\beta)$ and ${^{-1}}\mathcal{\hat{T}}^m_{\ell 2}(\nu, -\beta)$ up to second order in $\beta$ (see Appendix~\ref{app:KT_explicit}). 
However, for the following computations we only need the products $^{-1}\mathcal{\hat{T}}^0_{\ell 0}(\nu, -\beta)\,^{0}\mathcal{\hat{B}}^0_{0\ell'}(\nu, \beta)$ and $^{-1}\mathcal{\hat{T}}^m_{\ell 2}(\nu, -\beta)\,^{0}\mathcal{\hat{B}}^m_{2\ell'}(\nu, \beta)$ for which the expressions can be found in Appendix~\ref{app:KT_explicit_Thomson} up to second order in $\beta$. 

Once the required matrix elements are derived, it is rather easy to give the Thomson collision term in the lab frame, as presented in Appendix~\ref{eq:final_Doppler}. The expressions agree with previous works up to first order in $\beta$ \citep[e.g.,][]{Hu1994, Bartolo2007, Chluba2012}. At second order in $\beta$, only the terms related directly to the monopole, $n_{00}$, have been previously obtained \citep[see,][]{Hu1994, Bartolo2007, Chluba2012}. Here we extended to all terms $\propto \beta^2$. The computational demand for brute force approaches is quite high, but can be minimized substantially by using the boost operator approach, once the theoretical overhead is overcome. 

\section{Conclusions}
In this work we, considered the properties and calculation of the boost operator, generalized previous treatments, and showed that this operator can be easily obtained using the expressions for the (computationally simpler) aberration kernel [e.g., see Eq.~\eqref{eq:gen_kernel_op_ds_o_nu}]. We also generalized the differential equation approach for the aberration kernel to arbitrary Doppler weights [see Eq.~\eqref{eq:KD_gen}], eliminating the need to use raising and lowering operations in applications. This yielded a general operator differential equation for the boost operator [see Eq.~\eqref{eq:BD_gen}], concluding the more formal part of this work.

Given the wide range of problems in which the boost operator appears, we believe this work provides a framework for simplifying many cumbersome calculations. We considered the transformation of the low-$\ell$ CMB anisotropies (Sect.~\ref{sec:dipole_quadrupole}) and the kinematic corrections to the Thomson scattering process (Sect.~\ref{sec:kin_corrs_Thomson}) as work examples. However, the boost operator formalism may have additional applications for the modeling of the Sunyaev-Zeldovich kinematic corrections \citep{Challinor1998, Nozawa1998SZ, Chluba2005b, Chluba2012moments, Coulton2020} or searches for kinematically-induced anisotropies in the number counts of astrophysical sources \citep[e.g.,][]{Chluba2005b,Rubart2013, Dalang2022MNRAS, Wagenveld2025}. Cut-sky effects and boosting-induced corrections in detailed CMB analyses \citep[e.g.,][]{Jeong2014ab, Yasini2020, Coulton2020} may also be modeled more easily using this framework. 
Similarly, it may be possible to generalize the operator approach to time-dependent signals encountered in gravitational wave astronomy \citep{Bonvin2023, Cusin2024}. Finally, the boost operator approach can probably be used to model the Compton scattering process for moving ensembles of electrons, yielding alternative ways to derive the Kompaneets equation and its generalizations \citep[e.g.][]{Itoh98, Chluba2012}. However, we leave an exploration of these interesting questions to the future.

\small 

\bibliographystyle{mn2e}
\bibliography{Lit-2025}

\onecolumn

\begin{appendix}

\section{Second order Taylor coefficients for ${_s^d}\mathcal{\hat{B}}_{\ell \ell'}^{m}(\nu, \beta)$}
\label{app:B_coefficies}
By taking the rapidity derivative of Eq.~\eqref{eq:BD_gen}, we can obtain an expression for the second derivative. For convenience we introduce the operator
\bealf{
\label{eq:Ob_opp}
{^d}\mathcal{\hat{O}}_{\nu, \ell'}&=\frac{d-1+\mathcal{\hat{O}}_\nu}{\ell'},
}
which appears in each of the coefficients of the differential equation and simplifies recursive applications. This also means that 
$$\frac{1-d-\oOnu}{\ell'(\ell'+1)}=\frac{d-1+\oOnu}{\ell'}-\frac{d-1+\oOnu}{(\ell'+1)}={^d}\mathcal{\hat{O}}_{\nu, \ell'}-{^d}\mathcal{\hat{O}}_{\nu, \ell'+1}$$
For the second derivative, we then find 
\bealf{
\label{eq:BD_gen_result}
\partial^2_\eta \,{^d_s}\mathcal{\hat{B}}_{\ell\ell'}^{m}\big|_{\eta=0}  
\nonumber
&=
-(1+{^d}\mathcal{\hat{O}}_{\nu, \ell'+1})\,{_s}B^m_{\ell'+1}\,\partial_\eta{^d_s}\mathcal{\hat{B}}_{\ell \ell'+1}^{m}\big|_{\eta=0} 
-sm\,[{^d}\mathcal{\hat{O}}_{\nu, \ell'+1}-{^d}\mathcal{\hat{O}}_{\nu, \ell'}]\,\partial_\eta{^d_s}\mathcal{\hat{B}}_{\ell \ell'}^{m}\big|_{\eta=0} 
+(1-{^d}\mathcal{\hat{O}}_{\nu, \ell'})\,{_s}B^m_{\ell'}\,\partial_\eta{^d_s}\mathcal{\hat{B}}_{\ell \ell'-1}^{m}\big|_{\eta=0} 
\nonumber \\
&=(1+{^d}\mathcal{\hat{O}}_{\nu, \ell'+1})\,{_s}B^m_{\ell'+1}\,
\left\{
(1+{^d}\mathcal{\hat{O}}_{\nu, \ell'+2})\,{_s}B^m_{\ell'+2}\,\delta_{\ell \ell'+2}
+sm\,[{^d}\mathcal{\hat{O}}_{\nu, \ell'+2}-{^d}\mathcal{\hat{O}}_{\nu, \ell'+1}]\,\delta_{\ell \ell'+1}
-(1-{^d}\mathcal{\hat{O}}_{\nu, \ell'+1})\,{_s}B^m_{\ell'+1}\,\delta_{\ell \ell'}
\right\}
\nonumber \\
&\qquad 
+sm\,[{^d}\mathcal{\hat{O}}_{\nu, \ell'+1}-{^d}\mathcal{\hat{O}}_{\nu, \ell'}]
\left\{
(1+{^d}\mathcal{\hat{O}}_{\nu, \ell'+1})\,{_s}B^m_{\ell'+1}\,\,\delta_{\ell \ell'+1}
+sm\,[{^d}\mathcal{\hat{O}}_{\nu, \ell'+1}-{^d}\mathcal{\hat{O}}_{\nu, \ell'}]\,\delta_{\ell \ell'}
-(1-{^d}\mathcal{\hat{O}}_{\nu, \ell'})\,{_s}B^m_{\ell'}\,\,\delta_{\ell \ell'-1}
\right\}
\nonumber 
\\
&\qquad\qquad
-(1-{^d}\mathcal{\hat{O}}_{\nu, \ell'})\,{_s}B^m_{\ell'}\,
\left\{
(1+{^d}\mathcal{\hat{O}}_{\nu, \ell'})\,{_s}B^m_{\ell'}\,\,\delta_{\ell \ell'}
+sm\,[{^d}\mathcal{\hat{O}}_{\nu, \ell'}-{^d}\mathcal{\hat{O}}_{\nu, \ell'-1}]\,\delta_{\ell \ell'-1}
-(1-{^d}\mathcal{\hat{O}}_{\nu, \ell'-1})\,{_s}B^m_{\ell'-1}\,\,\delta_{\ell \ell'-2}
\right\}
\nonumber \\[2mm]
&=(1+{^d}\mathcal{\hat{O}}_{\nu, \ell'+1})(1+{^d}\mathcal{\hat{O}}_{\nu, \ell'+2})\,{_s}B^m_{\ell'+1}\,{_s}B^m_{\ell'+2} \,\delta_{\ell \ell'+2}
+sm\,(1+{^d}\mathcal{\hat{O}}_{\nu, \ell'+1})[{^d}\mathcal{\hat{O}}_{\nu, \ell'+2}-{^d}\mathcal{\hat{O}}_{\nu, \ell'}]\,{_s}B^m_{\ell'+1}\,\delta_{\ell \ell'+1}
\nonumber\\
&\qquad\qquad
-\left[
(1-{^d}\mathcal{\hat{O}}^2_{\nu, \ell'+1})\, ({_s}B^m_{\ell'+1})^2 
+s^2m^2\,[{^d}\mathcal{\hat{O}}_{\nu, \ell'+1}-{^d}\mathcal{\hat{O}}_{\nu, \ell'}]^2
+(1-{^d}\mathcal{\hat{O}}^2_{\nu, \ell'})\, ({_s}B^m_{\ell'})^2
\right]\delta_{\ell \ell'}
\nonumber \\
&\qquad\qquad\qquad
-sm\,(1-{^d}\mathcal{\hat{O}}_{\nu, \ell'})[{^d}\mathcal{\hat{O}}_{\nu, \ell'+1}-{^d}\mathcal{\hat{O}}_{\nu, \ell'-1}]\,{_s}B^m_{\ell'}\,\delta_{\ell \ell'-1}
+(1-{^d}\mathcal{\hat{O}}_{\nu, \ell'-1})(1-{^d}\mathcal{\hat{O}}_{\nu, \ell'})\,{_s}B^m_{\ell'-1}\,{_s}B^m_{\ell'} \,\delta_{\ell \ell'-2}.
}
Higher order terms can be obtained by repeatedly taking next order derivatives. At third order in $\beta$, corrections due to the transformation between $\eta$ and $\beta$ also have to be considered more carefully. Inserting the definition for ${^d}\mathcal{\hat{O}}_{\nu, \ell'}$ and simplifying the expressions then yields Eq.~\eqref{eq:d2Bdeta}.

\section{Explicit transformation of the CMB sky with a primordial dipole}
\label{app:dipole_transform}
To obtain an explicit expression for the CMB sky in the presence of a primordial dipole, we start with the expression for the transformed temperature variable. In the CMB restframe, we assume 
\bealf{
T(\vgh_{\rm c})
=T_0(1+\boldsymbol{\beta}_{\rm pr}\cdot \vgh_{\rm c}),
}
with the primordial dipole direction being determined by the vector $\boldsymbol{\beta}_{\rm pr}$. In the moving frame, we have to replace $\vgh_{\rm c}$ by \citep[e.g.,][]{Challinor2002}
\bealf{
\vgh_{\rm c}=\frac{\hat{\boldsymbol{\beta}}\cdot \vgh-\beta}{1-\boldsymbol{\beta}\cdot \vgh}\,\hat{\boldsymbol{\beta}}
+\frac{\vgh-(\hat{\boldsymbol{\beta}}\cdot \vgh)\,\hat{\boldsymbol{\beta}}}{\gamma(1-\boldsymbol{\beta}\cdot \vgh)}=\frac{\vgh-\gamma \,\boldsymbol{\beta}+(\gamma-1)(\hat{\boldsymbol{\beta}}\cdot \vgh)\,\hat{\boldsymbol{\beta}}}{\gamma(1-\boldsymbol{\beta}\cdot \vgh)}= \vgh-\boldsymbol{\beta}+(\boldsymbol{\beta}\cdot \vgh)\,\vgh+\mathcal{O}(\beta^2).
}
This then implies
\bealf{
\label{eq:Tprime_explicit}
\frac{T'(\vgh)}{T_0}
=\frac{1+\boldsymbol{\beta}_{\rm pr}\cdot \vgh_{\rm c}(\boldsymbol{\beta}, \vgh)}{\gamma (1-\boldsymbol{\beta}\cdot \vgh)}\approx 1+(\boldsymbol{\beta}+\boldsymbol{\beta}_{\rm pr})\cdot \vgh+(\boldsymbol{\beta}\cdot \vgh)^2+2(\boldsymbol{\beta}\cdot \vgh)\,(\boldsymbol{\beta}_{\rm pr}\cdot \vgh)-\frac{\beta^2}{2}-\boldsymbol{\beta}_{\rm pr}\cdot\boldsymbol{\beta}
\equiv 1+\boldsymbol{\beta}_{\rm t}\cdot \vgh+(\boldsymbol{\beta}_{\rm t}\cdot \vgh)^2-\frac{\beta^2_{\rm t}}{2}+\frac{\beta^2_{\rm pr}}{2}-(\boldsymbol{\beta}_{\rm pr}\cdot \vgh)^2,
}
where we introduced the total dipole velocity $\boldsymbol{\beta}_{\rm t}=\boldsymbol{\beta}+\boldsymbol{\beta}_{\rm pr}$. 
To simplify the subsequent computations, we use the addition theorem
\bealf{
\hat{\boldsymbol{n}}\cdot \vgh=
\frac{4\pi}{3}\sum_{m=-1}^1 Y^*_{1m}(\hat{\boldsymbol{n}}) \, Y_{1m}(\vgh)
\nonumber
}
to compute the harmonic coefficients when averaging over the direction $\vgh$. As an intermediate step, we may use
\bsub
\bealf{
[\hat{\boldsymbol{n}}_1\cdot \hat{\boldsymbol{n}}_2]_{\ell m}&=\sqrt{4\pi}\,\hat{\boldsymbol{n}}_1\cdot \hat{\boldsymbol{n}}_2\,\delta_{\ell 0}\,\delta_{m 0},
\qquad
[\hat{\boldsymbol{n}}\cdot \vgh]_{\ell m}=
\frac{4\pi}{3}\,Y^*_{1m}(\hat{\boldsymbol{n}}) \,\delta_{\ell 1},
\\
[(\hat{\boldsymbol{n}}_1\cdot \vgh)(\hat{\boldsymbol{n}}_2\cdot \vgh)]_{\ell m}&=
\left(\frac{4\pi}{3}\right)^2
\,\sum_{m'=-1}^1 \sum_{m''=-1}^1  Y_{1m'}(\hat{\boldsymbol{n}}_1)\,Y_{1m''}(\hat{\boldsymbol{n}}_2) \, \int Y_{\ell, m}(\vgh) Y_{1m'}(\vgh)Y_{1m''}(\vgh)\,\id^2 \vgh
\nonumber 
\\
&=\frac{(4\pi)^2}{3}
\,\sum_{m'=-1}^1 \sum_{m''=-1}^1  Y_{1 m'}(\hat{\boldsymbol{n}}_1)\,Y_{1 m''}(\hat{\boldsymbol{n}}_2) \,
\sqrt{\frac{(2\ell+1)}{4\pi}}
\threej{\ell}{1}{1}{0}{0}{0}
\threej{\ell}{1}{1}{m}{m'}{m''}
}
\esub
where $[X]_{\ell m}=\int Y^*_{\ell m}(\vgh) \,X(\vgh)\id \vgh$ is the spherical harmonic coefficient of $X(\vgh)$ and we used the $3J$-symbols to deal with the Gaunt integrals.
We therefore have the non-vanishing projections
\bsub
\label{eq:projection_integrals}
\bealf{
[(\hat{\boldsymbol{n}}_1\cdot \vgh)(\hat{\boldsymbol{n}}_2\cdot \vgh)]_{0 0}&=
\frac{\sqrt{4\pi}}{3} \times \frac{4\pi}{3}\sum_{m=-1}^1 Y^*_{1m}(\hat{\boldsymbol{n}}_1) \, Y_{1m}(\hat{\boldsymbol{n}}_2)
\equiv \frac{\sqrt{4\pi}}{3} \,\hat{\boldsymbol{n}}_1\cdot \hat{\boldsymbol{n}}_2
\\
[(\hat{\boldsymbol{n}}_1\cdot \vgh)(\hat{\boldsymbol{n}}_2\cdot \vgh)]_{2 m}&=\frac{8\pi}{3} \sqrt{\frac{2\pi}{15}}\;
\Bigg[
\sqrt{\frac{2}{3}}\, \left(Y_{10}(\hat{\boldsymbol{n}}_1) \, Y_{10}(\hat{\boldsymbol{n}}_2)+
\frac{Y_{11}(\hat{\boldsymbol{n}}_1) \, Y_{1-1}(\hat{\boldsymbol{n}}_2)+Y_{1-1}(\hat{\boldsymbol{n}}_1) \, Y_{11}(\hat{\boldsymbol{n}}_2)}{2}\right)\delta_{m0}
\nonumber\\
&\qquad\qquad\qquad\qquad\qquad\qquad\qquad
+
\frac{Y^*_{1m}(\hat{\boldsymbol{n}}_1) \, Y_{10}(\hat{\boldsymbol{n}}_2)+Y_{10}(\hat{\boldsymbol{n}}_1) \, Y^*_{1m}(\hat{\boldsymbol{n}}_2)}{\sqrt{2}} \,\delta_{|m|1}
+Y_{11}(\hat{\boldsymbol{n}}_1) \, Y_{11}(\hat{\boldsymbol{n}}_1) \,\delta_{m,-2}
+Y^*_{11}(\hat{\boldsymbol{n}}_1) \, Y^*_{11}(\hat{\boldsymbol{n}}_1) \,\delta_{m,2}
\Bigg]
\\
[(\hat{\boldsymbol{n}}\cdot \vgh)^2]_{2 m}&=\frac{8\pi}{3} \sqrt{\frac{2\pi}{15}}\;
\Bigg[
\sqrt{\frac{2}{3}}\, \left(Y^2_{10}(\hat{\boldsymbol{n}})+
Y_{11}(\hat{\boldsymbol{n}}) \, Y_{1-1}(\hat{\boldsymbol{n}})\right)\delta_{m0}
+
\sqrt{2}\, Y^*_{1m}(\hat{\boldsymbol{n}}) \, Y_{10}(\hat{\boldsymbol{n}}) \,\delta_{|m|1}
+Y^2_{11}(\hat{\boldsymbol{n}}) \,\delta_{m,-2}
+[Y^*_{11}(\hat{\boldsymbol{n}})]^2 \,\delta_{m,2}
\Bigg]
\equiv \frac{8\pi}{15} Y^*_{2m}(\hat{\boldsymbol{n}})
}
\esub
With these relations, it is then easy to compute the results for the spectra of the low-$\ell$ multipole.

Using Eq.~\eqref{eq:projection_integrals} with Eq.~\eqref{eq:Tprime_explicit}, we find the average temperature as \citep[see also][]{Chluba2011ab}
\bealf{
\label{eq:av_T_on_temperature}
T'_0=\int T'(\vgh) \frac{\id^2\vgh}{4\pi} 
\approx T_0
\left[
1-\frac{\beta^2}{6}-\frac{\boldsymbol{\beta}\cdot\boldsymbol{\beta}_{\rm pr}}{3}
\right]
\equiv T_0
\left[
1-\frac{\beta_{\rm t}^2}{6}+\frac{\beta_{\rm pr}^2}{6}
\right].
}
In the presence of observer motion the primordial CMB dipole changes the average CMB blackbody temperature at second order in the small quantities \citep{Chluba2011ab}. If the total dipole, $\boldsymbol{\beta}_{\rm t}=\boldsymbol{\beta}+\boldsymbol{\beta}_{\rm pr}$, was purely motion-induced, we would only have the first contribution, $\propto(\boldsymbol{\beta}+\boldsymbol{\beta}_{\rm pr})^2/6$. The second term corrects for the fact that $\boldsymbol{\beta}_{\rm pr}$ does not contribute to the motion.

To obtain the final expressions for the transformed photon occupation number at second order in the small variables, we use a second order expansion in the temperature perturbations
\bealf{
n'(\nu, \vgh)\approx \frac{1}{\expf{x}-1}+\left[\Delta_T(\vgh)+\Delta^2_T(\vgh)\right]\,G(x)+\frac{\Delta^2_T(\vgh)}{2}\,Y(x),
}
with $\Delta_T(\vgh)=T'(\vgh)/T_0-1$ and $T'(\vgh)/T_0$ from Eq.~\eqref{eq:Tprime_explicit}, and $\Delta^2_T(\vgh)$ given by
\bealf{
\Delta^2_T(\vgh) \approx [(\boldsymbol{\beta}+\boldsymbol{\beta}_{\rm pr})\cdot \vgh]^2=(\boldsymbol{\beta}\cdot\vgh)^2
+(\boldsymbol{\beta}_{\rm pr}\cdot\vgh)^2+2(\boldsymbol{\beta}\cdot\vgh)(\boldsymbol{\beta}_{\rm pr}\cdot\vgh).
}
This second order contribution is due to the superposition of blackbodies and can be avoided by working with thermodynamic temperature variables.
From the superposition effect, the $y$-parameter of the transformed spectrum is then $y_{\rm t} = [\Delta^2_T(\vgh)]_{0}/2= [\Delta^2_T(\vgh)]_{00}/[2\sqrt{4\pi}]=\beta^2_{\rm t}/6$ and the extra contribution to the average temperature is $\Delta T/T_0\approx \beta^2_{\rm t}/3$, with Eq.~\eqref{eq:av_T_on_temperature} yielding the result in Eq.~\eqref{eq:transformed_sky_both_monopole}. The transformed dipole is given by $n'_{1m}(\nu)\approx [\Delta_T(\vgh)]_{1m} \,G(x)=[(\boldsymbol{\beta}+\boldsymbol{\beta}_{\rm pr})\cdot \vgh]_{1m}\,G(x)$. For the quadrupole we have two terms,
\bealf{
\label{eq:Quadrupole_general}
n'_{2m}(\nu)\approx \left\{[(\boldsymbol{\beta}\cdot \vgh)^2]_{2m} 
+2 [(\boldsymbol{\beta}\cdot \vgh)(\boldsymbol{\beta}_{\rm pr}\cdot \vgh)]_{2m} + [(\boldsymbol{\beta}_{\rm t}\cdot \vgh)^2]_{2m}\right\}\,G(x)+\frac{[(\boldsymbol{\beta}_{\rm t}\cdot \vgh)^2]_{2m}}{2}\,Y(x).
}
where the terms $\propto [\Delta^2_T(\vgh)]_{2m}=[(\boldsymbol{\beta}_{\rm t}\cdot \vgh)^2]_{2m}$ stem from the superposition blackbody spectra. Note that as as simplification, we can also write $$[(\boldsymbol{\beta}\cdot \vgh)^2]_{2m} 
+2 [(\boldsymbol{\beta}\cdot \vgh)(\boldsymbol{\beta}_{\rm pr}\cdot \vgh)]_{2m} + [(\boldsymbol{\beta}_{\rm t}\cdot \vgh)^2]_{2m}=2[(\boldsymbol{\beta}_{\rm t}\cdot \vgh)^2]_{2m}-[(\boldsymbol{\beta}_{\rm pr}\cdot \vgh)^2]_{2m}.$$
This then explains all terms in Eq.~\eqref{eq:transformed_sky_both}.

\section{Explicit expressions for ${^0}\mathcal{\hat{B}}^m_{\ell\ell'}(\beta)$ and $^{-1}\mathcal{\hat{T}}^m_{\ell\ell'}(\nu, -\beta)$ ($s=0$)}
\label{app:KT_explicit}
To compute the Thomson terms we require 
\bsub
\bealf{
^0\mathcal{\hat{B}}^m_{0\ell}(\nu, \beta)
&\approx 
\left[1+\frac{1}{6}\oOnu\,\beta^2+\frac{1}{6}\oDnu\,\beta^2\right]\,\delta_{0\ell}\delta_{m0}+\Bigg[2-\oOnu\Bigg]\,\frac{\beta}{\sqrt{3}} \,\delta_{1\ell}\delta_{m0}
+\left[3-\oOnu +\frac{1}{2}\oDnu\right]\,\frac{2\beta^2}{3\sqrt{5}} \,\delta_{2\ell}\delta_{m0}
\\
^0\mathcal{\hat{B}}^m_{2\ell}(\nu, \beta)
&\approx 
 \left[4\oOnu + \oDnu \right]\,\frac{\beta^2}{3\sqrt{5}}\,\delta_{0\ell}\delta_{m0}
-C^m_2\Bigg[1+\oOnu\Bigg]\,\beta \,\delta_{1\ell}+
\left[1
-\left\{(C^m_3)^2\left(8-\oOnu-\oDnu \right)
+(C^m_2)^2\left(3-\oOnu-\oDnu \right)
\right\}\,\frac{\beta^2}{2}
\right]\delta_{2\ell}
\nonumber
\\
&\qquad\qquad\qquad\qquad\qquad\qquad
+C^m_3\Bigg[4-\oOnu\Bigg]\,\beta \,\delta_{3\ell}
+C^m_3 C^m_4\left[10-3\oOnu+\frac{1}{2}\oDnu\right]\beta^2 \,\delta_{4\ell}
\\[2mm]
^{-1}\mathcal{\hat{T}}^m_{\ell0}(\nu, -\beta)
&\approx 
\left[1-\frac{1}{6}\oOnu\,\beta^2+\frac{1}{6}\oDnu\,\beta^2 \right]\,\delta_{\ell 0}\delta_{m0}
-\Bigg[1-\oOnu\Bigg]\,\frac{\beta}{\sqrt{3}} \,\delta_{\ell 1}\delta_{m0}
+\left[\oOnu+\frac{1}{2}\oDnu\right]\,\frac{2\beta^2}{3\sqrt{5}} \,\delta_{\ell 2}\delta_{m0}
\\
^{-1}\mathcal{\hat{T}}^m_{\ell2}(\nu, -\beta)
&\approx 
\left[6-2\oOnu+\frac{1}{2}\oDnu\right]\frac{2\beta^2}{3\sqrt{5}}\,\delta_{\ell0}\delta_{m0}
-C^m_2\Bigg[4-\oOnu\Bigg]\,\beta \,\delta_{\ell1}
+
\left[1
-\left\{(C^m_3)^2\left(12+\oOnu-\oDnu \right)
-(C^m_2)^2\left(3-\oOnu+\oDnu \right)
\right\}\,\frac{\beta^2}{2}
\right]\delta_{2\ell}
\nonumber
\\
&\qquad\qquad\qquad\qquad\qquad\qquad
+C^m_3\Bigg[1+\oOnu\Bigg]\,\beta \,\delta_{\ell 3}
+C^m_3 C^m_4\Bigg[1+3\oOnu+\frac{1}{2}\oDnu\Bigg]\,\beta^2 \,\delta_{\ell 4}.
}
\esub
Note that $C^m_\ell=0$ for $|m|>\ell$.

\section{Explicit expressions for $^{-1}\mathcal{\hat{T}}^m_{\ell \ell''}(\nu, -\beta)\,^{0}\mathcal{\hat{B}}^m_{\ell''\ell'}(\nu, \beta)$ ($s=0$ and $\ell''=0$ and $\ell''=2$)}
\label{app:KT_explicit_Thomson}
For the transformation back into the lab frame, we require
\bsub
\label{eq:all_kernels}
\bealf{
^{-1}\mathcal{\hat{T}}^m_{\ell 0}(\nu, -\beta)\,^{0}\mathcal{\hat{B}}^m_{0\ell'}
(\nu, \beta)&
\approx 
\Bigg\{
\delta_{\ell 0}\delta_{0\ell'}
+
\Big[
 \left(2-\oOnu\right)\delta_{\ell 0}\delta_{1\ell'}
-\left(1-\oOnu\right)\delta_{\ell 1}\delta_{0\ell'}
\Big]\frac{\beta}{\sqrt{3}}
+
\Big[
\oDnu\,
\delta_{\ell 0}\delta_{0\ell'}
-
\left(
2+\oDnu
\right)\delta_{\ell 1}\delta_{1\ell'}
\Big]\,\frac{\beta^2}{3}
\\
&\qquad\qquad\qquad
+
\left[
\left(6-2\oOnu+\oDnu\right)
\delta_{\ell 0}\delta_{2\ell'}
+
\left(2\oOnu+\oDnu\right)\delta_{\ell 2}\delta_{0\ell'}
\right]\frac{\beta^2}{3\sqrt{5}}
\Bigg\}\,\delta_{m0}
\nonumber \\[1mm]
^{-1}\mathcal{\hat{T}}^m_{\ell 2}(\nu, -\beta)\,^{0}\mathcal{\hat{B}}^m_{2\ell'}(\nu, \beta)&
\approx 
\delta_{\ell 2}\delta_{2\ell'}-C^m_2
\bigg[
\left(4-\oOnu\right)\delta_{\ell 1}\delta_{2\ell'}
+
\left(1+\oOnu\right)\delta_{\ell 2}\delta_{1\ell'}
\bigg]\,\beta
+C^m_3
\bigg[
\left(4-\oOnu\right)\delta_{\ell 2}\delta_{3\ell'}
+
 \left(1+\oOnu\right)\delta_{\ell 3}\delta_{2\ell'}
\bigg]\,\beta
\nonumber\\
&\!\!\!\!\!\!\!\!\!\!\!\!\!\!\!\!\!\!
+ C^m_1C^m_2 
\left[
\left(2\oOnu+\frac{1}{2}\oDnu\right)\,\delta_{\ell 2}\delta_{0\ell'}
+\left(6-2\oOnu+\frac{1}{2}\oDnu\right) \delta_{\ell 0}\delta_{2\ell'}
\right]\,\beta^2
+(C^m_2)^2\left(4-\oDnu\right)\beta^2\,\delta_{\ell 1}\delta_{1\ell'}
\nonumber\\
&\!\!\!\!\!\!\!\!\!
+\left[
(C^m_2)^2\oDnu
-(C^m_3)^2\left(10-\oDnu \right)
\right]\beta^2\,\delta_{\ell 2}\delta_{2\ell'}
-C^m_2 C^m_3\left[\left(1+5\oOnu+\oDnu\right)\,\delta_{\ell 3}\delta_{1\ell'}
+\left(16-5\oOnu+\oDnu\right)\,\delta_{\ell 1}\delta_{3\ell'}\right]\beta^2
\nonumber\\
&
+(C^m_3)^2\left(4-\oDnu\right)\beta^2\,\delta_{\ell 3}\delta_{3\ell'}
+C^m_3 C^m_4\left[
\left(10-3\oOnu+\frac{1}{2}\oDnu\right) \,\delta_{\ell 2}\delta_{4\ell'} 
+
\left(1+3\oOnu+\frac{1}{2}\oDnu\right) \,\delta_{\ell 4}\delta_{2\ell'}\,\right]\beta^2 
}
\esub
Here we used $\oDnu=4\nu\partial_\nu+\nu^2\partial^2_\nu$.

\section{Final Thomson collision term at second order in $\beta=\varv/c$}
\label{eq:final_Doppler}
Inserting Eq.~\eqref{eq:all_kernels} into Eq.~\eqref{eq:dn_transformed} and collecting terms in orders of $\beta$ we then finally have
\bsub
\bealf{
\left.\frac{\id  n_{00}(\nu)}{ \id \tau}\right|_{\rm T}
&\!=\!\frac{\beta}{\sqrt{3}}\left(3-\oOnu\right)\,n_{10}
+\frac{\beta^2}{3}\oDnu\,n_{00}
+\frac{\beta^2}{\sqrt{5}}\left[\frac{4}{5}\,(3-\oOnu)+\frac{11}{30}\oDnu \right]n_{20}
\\[2mm]
\left.\frac{\id  n_{10}(\nu)}{ \id \tau}\right|_{\rm T}
&\!=\!-n_{10}
+\frac{\beta}{\sqrt{3}}\oOnu n_{00}
+\frac{\beta}{5\sqrt{15}}(6+\oOnu)n_{20}
-\frac{\beta^2}{25}\,\left[14+9\oDnu\right]n_{10}
-\frac{\beta^2}{25}\sqrt{\frac{3}{7}}\,\left[16-5\oOnu+\oDnu \right]n_{30}
\\
\left.\frac{\id  n_{1\pm1}(\nu)}{ \id \tau}\right|_{\rm T}
&\!=\!-n_{1\pm1}+\frac{\beta}{10\sqrt{5}}\,(6+\oOnu)n_{2\pm1}
+\frac{\beta^2}{50}\,\left[4-\oDnu\right]n_{1\pm1}
-\frac{\beta^2}{25}\sqrt{\frac{2}{7}}\,\left[16-5\oOnu+\oDnu \right]n_{3\pm1}
\\[2mm]
\left.\frac{\id  n_{20}(\nu)}{ \id \tau}\right|_{\rm T}
&\!=\!-\frac{9}{10}n_{20}
+\frac{\beta}{5\sqrt{15}}\left(9-\oOnu\right) n_{10}
+\frac{3\beta}{10\sqrt{35}}\left(14-\oOnu\right) n_{30}
+\frac{\beta^2}{\sqrt{5}}\left[\frac{4}{5}\oOnu+\frac{11}{30}\oDnu \right]n_{00}
-\frac{\beta^2}{3}\left[\frac{27}{35}-\frac{11}{70}\oDnu\right]n_{20}
+\frac{\beta^2}{35\sqrt{5}}\left[20-6\oOnu+\oDnu\right]n_{40}
\\
\left.\frac{\id  n_{2\pm1}(\nu)}{ \id \tau}\right|_{\rm T}
&\!=\!-\frac{9}{10}n_{2\pm1}
+\frac{\beta}{10\sqrt{5}}\,\left(9-\oOnu\right)\,n_{1\pm1}
+\frac{\beta}{5}\sqrt{\frac{2}{35}}\,\left(14-\oOnu\right)\,n_{3\pm1}
-\beta^2\,\left[\frac{8}{35}-\frac{3}{70}\oDnu\right]n_{2\pm1}
+\frac{\beta^2}{35\sqrt{6}}\,\left[20-6\oOnu+\oDnu\right]\,n_{4\pm1}
\\
\left.\frac{\id  n_{2\pm2}(\nu)}{ \id \tau}\right|_{\rm T}
&\!=\!-\frac{9}{10}n_{2\pm2}
+\frac{\beta}{10\sqrt{7}}\,\left(14-\oOnu\right)\,n_{3\pm2}
-\frac{\beta^2}{70}\,\left[10-\oDnu\right]n_{2\pm2}
+\frac{\beta^2}{70\sqrt{3}}\,\left[20-6\oOnu+\oDnu\right]\,n_{4\pm2}
\\[2mm] 
\left.\frac{\id  n_{30}(\nu)}{ \id \tau}\right|_{\rm T}
&\!=\!-n_{30}
+\frac{3\beta}{10\sqrt{35}}\,\left(11+\oOnu\right)\,n_{20}
+\frac{4\beta}{3\sqrt{7}}\,n_{40}
-\frac{\beta^2}{25}\sqrt{\frac{3}{7}}\,\left[1+5\oOnu+\oDnu \right]n_{10}
+\frac{9\beta^2}{350}\,\left[4-\oDnu \right]n_{30}
\\ 
\left.\frac{\id  n_{3\pm1}(\nu)}{ \id \tau}\right|_{\rm T}
&\!=\!-n_{3\pm1}\!
+\frac{\beta}{5}\!\sqrt{\frac{2}{35}}\left(11+\oOnu\right) n_{2\pm1}
+ \beta \!\sqrt{\frac{5}{21}} n_{4\pm1}
-\frac{\beta^2}{25}\sqrt{\frac{2}{7}}\,\left[1+5\oOnu+\oDnu \right]n_{1\pm1}
+\frac{4\beta^2}{175}\,\left[4-\oDnu \right]n_{3\pm1}
\\ 
\left.\frac{\id  n_{3\pm2}(\nu)}{ \id \tau}\right|_{\rm T}
&\!=\!-n_{3\pm2}
+\frac{\beta}{10\sqrt{7}}\,\left(11+\oOnu\right)\,n_{2\pm2}
+ \beta \sqrt{\frac{4}{21}}\,n_{4\pm2}
+\frac{\beta^2}{70}\,\left[4-\oDnu \right]n_{3\pm2}
\\ 
\left.\frac{\id  n_{3\pm3}(\nu)}{ \id \tau}\right|_{\rm T}
&\!=\!-n_{3\pm3}
+ \frac{\beta}{3}\,n_{4\pm3}
\\[2mm] 
\left.\frac{\id  n_{40}(\nu)}{ \id \tau}\right|_{\rm T}
&\!=\!-n_{40}+\frac{4\beta}{3\sqrt{7}}\,n_{30}+\frac{5\beta}{3\sqrt{11}}\,n_{50}
+\frac{2\beta^2}{35\sqrt{5}}\,\left[1+3\oOnu+\frac{\oDnu}{2} \right]n_{20}
\\ 
\left.\frac{\id  n_{4\pm1}(\nu)}{ \id \tau}\right|_{\rm T}
&\!=\!-n_{4\pm1}\!+\beta\sqrt{\frac{5}{21}}\,n_{3\pm1}+\beta\sqrt{\frac{8}{21}}\,n_{5\pm1}
+\frac{2\beta^2}{35\sqrt{6}}\,\left[1+3\oOnu+\frac{\oDnu}{2} \right]n_{2\pm1}
\\ 
\left.\frac{\id  n_{4\pm2}(\nu)}{ \id \tau}\right|_{\rm T}
&\!=\!-n_{4\pm2}\!+\beta\sqrt{\frac{4}{21}}\,n_{3\pm2}+\beta\sqrt{\frac{7}{33}}\,n_{5\pm2}
+\frac{\beta^2}{35\sqrt{3}}\,\left[1+3\oOnu+\frac{\oDnu}{2} \right]n_{2\pm2}
\\ 
\left.\frac{\id  n_{4\pm3}(\nu)}{ \id \tau}\right|_{\rm T}
&\!=\!-n_{4\pm3}\!+\frac{\beta}{3}\,n_{3\pm3}+\frac{4\beta}{3\sqrt{11}}\,n_{5\pm3}
\\ 
\left.\frac{\id  n_{4\pm4}(\nu)}{ \id \tau}\right|_{\rm T}
&\!=\!-n_{4\pm4}\!+\frac{\beta}{\sqrt{11}}\,n_{5\pm4}
\\[2mm]
\left.\frac{\id  n_{\ell m}(\nu)}{ \id \tau}\right|_{\rm T}
&\!=\! -n_{\ell m} + \beta \left[C_{\ell+1}^m n_{\ell+1 m} +C_{\ell}^m n_{\ell-1 m}\right]
\;\,\text{$(\ell>4)$}
}
\esub
up to second order in $\beta$. We confirmed all these terms by explicit integration.

\end{appendix}

\end{document}